\documentclass[%
 aip,
 amsmath,amssymb,
preprint,%
]{revtex4-1}

\usepackage{graphicx}
\usepackage{dcolumn}
\usepackage{bm}

\usepackage[utf8]{inputenc}
\usepackage[T1]{fontenc}
\usepackage{mathptmx}
\usepackage{etoolbox}
\usepackage{xcolor}
\usepackage{subfig}

\makeatletter
\def\@email#1#2{%
 \endgroup
 \patchcmd{\titleblock@produce}
  {\frontmatter@RRAPformat}
  {\frontmatter@RRAPformat{\produce@RRAP{*#1\href{mailto:#2}{#2}}}\frontmatter@RRAPformat}
  {}{}
}%
\makeatother
\begin{document}
\raggedbottom

\title{Learning thermodynamically constrained equations of state with uncertainty}
\author{Himanshu Sharma}
\affiliation{ 
Dept. of Civil and Systems Engineering, Johns Hopkins University, Baltimore, MD, 21212, USA
}


\author{Jim A Gaffney}
\affiliation{Lawrence Livermore National Laboratory, 7000 East Ave, Livermore, CA 94550, USA
}%

\author{Dimitrios Tsapetis}%
\affiliation{ 
Dept. of Civil and Systems Engineering, Johns Hopkins University, Baltimore, MD, 21212, USA
}%

\author{Michael D.\ Shields}%
 \email{michael.shields@jhu.edu}
\affiliation{ 
Dept. of Civil and Systems Engineering, Johns Hopkins University, Baltimore, MD, 21212, USA
}%

\date{\today}

\begin{abstract}

Numerical simulations of high energy-density experiments require equation of state (EOS) models that relate a material's thermodynamic state variables -- specifically pressure, volume/density, energy, and temperature. EOS models are typically constructed using a semi-empirical parametric methodology, which assumes a physics-informed functional form with many tunable parameters calibrated using experimental/simulation data. Since there are inherent uncertainties in the calibration data (parametric uncertainty) and the assumed functional EOS form (model uncertainty), it is essential to perform uncertainty quantification (UQ) to improve confidence in the EOS predictions. Model uncertainty is challenging for UQ studies since it requires exploring the space of all possible physically consistent functional forms. Thus, it is often neglected in favor of parametric uncertainty, which is easier to quantify without violating thermodynamic laws. This work presents a data-driven machine learning approach to constructing EOS models that naturally captures model uncertainty while satisfying the necessary thermodynamic consistency and stability constraints. We propose a novel framework based on physics-informed Gaussian process regression (GPR) that automatically captures total uncertainty in the EOS and can be jointly trained on both simulation and experimental data sources. A GPR model for the shock Hugoniot is derived and its uncertainties are quantified using the proposed framework. We apply the proposed model to learn the EOS for the diamond solid state of carbon, using both density functional theory data and experimental shock Hugoniot data to train the model and show that the prediction uncertainty is reduced by considering the thermodynamic constraints.  
\end{abstract}

\maketitle

\section{\label{sec:level1}Introduction \protect\\ }

Hydrodynamics simulations, which are widely used to predict and understand the evolution of experiments in high energy density physics, inertial confinement fusion, laboratory astrophysics, and geophysics, are underpinned by equation of state (EOS) models which are needed to relate the thermodynamic state variables of the materials of interest\cite{gaffney2018review}. The accuracy and precision of the EOS, and the development of methods to quantify uncertainty in EOS models, is therefore a crucial concern. This is a challenging task that will require novel methods to complete.

EOS models are typically constructed using semi-empirical functions where the functional form is motivated by the physics, and the parameters are calibrated using a complex combination of experimental and first-principles simulation data from a variety of sources. Once calibrated, the EOS model can be used to interpolate and extrapolate over the wide range of input states needed in hydrodynamic simulations. The semi-empirical approach is subject to two sources of uncertainty: uncertainty in the values of the parameters in the EOS model (parameter uncertainty) and uncertainty in the form of the EOS model itself (model uncertainty). Both sources must be quantified simultaneously to give a complete picture of the total EOS uncertainty. While parametric uncertainty in the EOS has been addressed in several recent works, model uncertainty remains a significant challenge. In this work, we describe a new machine learning based approach to UQ which accounts for \emph{all} sources of uncertainty and provides an analytical framework for combining heterogeneous data sources into a single, uncertainty-aware EOS model.

Our new approach uses Gaussian process (GP) regression\cite{williams2006gaussian} to construct a data-driven EOS model that automatically satisfies all thermodynamic constraints. The resulting model provides pointwise predictions in the thermodynamic state space that include both model and data uncertainty. Incorporating thermodynamic constraints ensures that predictions satisfy the underlying physics across the entire domain, thereby avoiding pathologies that lead to the failure of downstream tasks like hydrodynamics modeling. Next, we derive a GP model for the shock Hugoniot directly from the uncertain EOS. This allows us to derive a novel unified approach enabling the model to be trained from first-principles simulation data, various experimental data sources, or both. We apply the proposed method to EOS modeling for the diamond phase of carbon, for which first-principles simulation data were first used to train the model. Then, experimental Hugoniot data are integrated into the unified GP EOS to create a jointly-trained uncertain EOS. The proposed model provides a powerful yet flexible non-parametric EOS that can be learned directly from heterogeneous data, obeys the important thermodynamic principles, and quantifies uncertainties that stem from noisy and sparse data from disparate sources.

\section{EOS Models with Uncertainty: Motivation \& Methods}

\subsection{\label{sec:prior_work}Relevant Prior Work \protect\\ }

In the standard setting, EOS parameter calibration depends on individual modelers who leverage domain knowledge and expertise to align EOS predictions with given experimental and simulation data. Recently, it has been common to pose the calibration process as an optimization problem to solve for the best parameters that give the least prediction error compared to available data\cite{forte2018multi,bergh2018optimization,cox2015fitting,myint2021minimization}. This approach can naturally be extended to capture parametric uncertainty by considering uncertainties in the calibration data sets. \citet{ali2020development} proposed a method that considers small perturbations in experimental data to calibrate model parameters using an optimization routine and propagate the experimental uncertainties using Monte Carlo simulation through the EOS models. \citet{brown2018estimating} apply Bayesian model calibration to estimate parameters using dynamic material properties experiments under extreme conditions. \citet{lindquist2022uncertainty} proposed a Bayesian framework to perform UQ of a multi-phase EOS model for carbon by accounting for calibration data uncertainty and yielding an ensemble of model parameters set. \citet{walters2018bayesian} used a Bayesian statistical approach to quantify parametric uncertainty by coupling hydrocode simulations and velocimetry measurements of a series of plate impact experiments of an aluminum alloy. 
\citet{robinson2013fundamental} have quantified uncertainties in the EOS stemming from the measurement noise in the experimental data and propagated these uncertainties for hydrocode simulations. They have emphasized the crucial need to integrate model uncertainty into consideration, identifying it as a significant factor.

Quantifying model form uncertainty, on the other hand, is more challenging since it requires exploring the infinite-dimensional space of possible functional forms that are thermodynamically constrained. Nonetheless, some work has been done to explore model uncertainty \cite{kamga2014optimal, gaffney2022constraining}. For example, \citet{kamga2014optimal} have performed UQ in a single model by exploring discrepancies in legacy experimental data. \citet{gaffney2022constraining} used GP regression to capture model uncertainty in the EOS of B$_4$C, accounting for the thermodynamic consistency constraint by explicitly modeling the free energy. They showed that the constraint reduces model uncertainty in the EOS by limiting the space of functions that can be fit to first-principles simulations. However, their model ignores the important thermodynamics stability constraints, which ensure that the specific heat and isothermal compressibility remains positive, and that will quickly cause hydrodynamic simulations to fail when violated.

\subsection{Uncertainty in Parametric EOS}

An EOS model is a semi-empirical equation that relates a set of state variables in a material such as temperature $T$, mass density $\rho$ (or volume $V$), pressure $P$, internal energy $E$, entropy $S$, etc. The standard process of building EOS models 
is to leverage expert knowledge of the material state under different conditions and assume a functional form that obeys the laws of thermodynamics. These often involve many parameters that must be carefully calibrated 
using a combination of experimental results 
and first-principles simulations. A generic EOS model may be written as,
\begin{equation}
\mathbf{F}_\alpha(\boldsymbol{\Theta})=0
\end{equation} 
where $\mathbf{F}$ is a vector function relating a set of state variables, $\alpha$ are the set of parameters unique to the assumed EOS model and $\boldsymbol{\Theta}$ are the set of state variables (e.g.\ $\boldsymbol{\Theta}=\{P,V,T,E\}$) that the EOS relates. In hydrodynamic simulations, for example, it is common to express the EOS in terms of the volume $V$ and temperature $T$ in the following form:
\begin{equation}
{\{P, E\}} = \mathbf{F}_\alpha(T,V)
\label{eqn:EOS_Hyro}
\end{equation}
Once the parameters $\alpha$ are learned, the EOS model can be utilized to predict the desired material state in the thermodynamics phase space. 

To learn these parameters with uncertainty, we can solve the requisite inverse problem in a Bayesian setting. Here, we can determine the distribution of the parameters $\alpha$ conditioned on the observed data $\bm{d}$ as:
\begin{equation}
    p(\alpha | \bm{d}) = \dfrac{p(\bm{d}|\alpha) p(\alpha)}{p(\bm{d})}
    \label{eqn:Bayes}
\end{equation}
where $p(\bm{d}|\alpha)$ is the likelihood function, $p(\alpha)$ is the prior distribution reflecting our existing knowledge of the parameters, and $p(\bm{d})$ is the evidence that serves as a normalization and does not need to be computed in the application for parameter estimation. In a general setting, this Bayesian inference problem is solved indirectly by drawing samples from $p(\alpha | \bm{d})$ using various Markov Chain Monte Carlo (MCMC) methods. As we'll see, this Bayesian inference process can be difficult when data are limited and/or the parameter vector $\alpha$ is very high-dimensional. 

UQ for existing parametric EOS models is further limited by the prescribed form of these models. Although often derived from physical principles, these models are nonetheless built upon assumptions, simplifications, and approximations of known physics, while neglecting physics that are poorly understood. Consequently, these models may be very accurate in certain regimes (e.g.\ of $T$ and $P$) and inadequate in others. The resulting uncertainty in these predictions is referred to as model-form uncertainty and cannot be accounted for in existing parametric models. In certain cases, competing parametric models can be compared and selected using Bayesian model selection. However, this requires the computation of the evidence term (denominator in Eq.\ \eqref{eqn:Bayes}), which poses significant practical challenges. 
Model-form uncertainty, combined with parametric uncertainty, results in a range of outputs for a fixed input state, thus yielding an ensemble of valid EOS models. 
Hence, it is also necessary to quantify these model-form uncertainties in a rigorous UQ framework to enhance our confidence in EOS predictions. Several approaches exist in the literature to address the model form uncertainties in Bayesian model calibration. These include the Kennedy and O’Hagan (KOH) approach\cite{kennedy2001bayesian}, the hierarchical Bayesian approach\cite{behmanesh2015hierarchical}, and the Bayesian model averaging approach\cite{park2011quantifying,hoeting1999bayesian}, among others. Specifically, in the KOH approach, a discrepancy model is introduced into the mathematical model to address model form uncertainty. However, these methodologies have not been extensively explored in the context of EOS models, partly due to challenges associated with their implementation and computational complexity. Our approach is complementary to these previous works as we can use it to introduce physics constraints into the existing uncertainty frameworks as appropriate for a specific problem; for example, we could add a physically constrained GP correction term to the physics-based parametric models in the manner of KOH. 
Significant research efforts have been made in quantifying parametric uncertainty; however, model-form UQ for equations of state has not received adequate attention, and only a few publications are available in the literature \cite{kamga2014optimal,gaffney2022constraining}. The biggest hurdle in quantifying model-form uncertainty is enumerating all the potential mappings that could form physics-consistent EOSs. In this work, we develop a framework using Gaussian process regression that automatically explores the range of physics-consistent EOS models to capture both sources of uncertainty. The proposed method is non-parametric and data-driven, yet satisfies both thermodynamic consistency and stability constraints. 

\subsection{Parametric EOS with Uncertainty: An Illustration}
\begin{figure}[h!]
   \centering 
    \begin{tabular}{cc}
\includegraphics[width=0.6\textwidth]{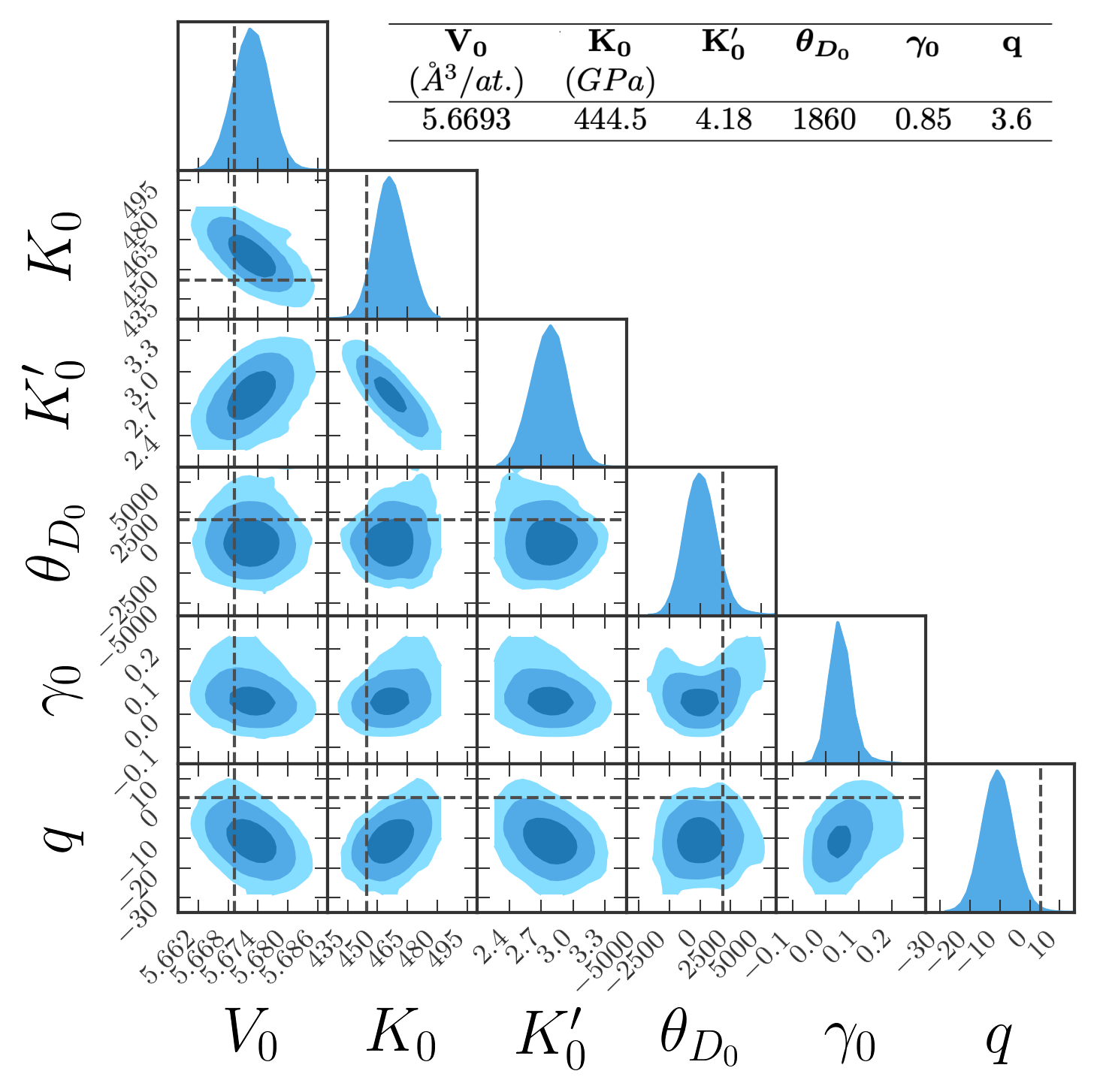}
\end{tabular}
\caption{Joint probability distribution function from Bayesian calibration of the Mie-Gr{\"u}neisen-Debye EOS model for diamond phase carbon. Inset table: Deterministic parameters from Dawaele et al.\cite{dewaele2008high}}
\label{fig:MGD_Model}
\end{figure}

An example of a parametric equation of state is the Mie-Gr{\"u}neisen-Debye model \cite{fei2007toward,fei2007toward} for single phase diamond. This model, which has been widely used for modeling materials (e.g.\ carbon and neon) in high temperature and high pressure environments \cite{} has the following form:
\begin{equation}
P(V,T) = P_V(V,T=0) + P_{TH\text{ Debye}}(V,T)
\end{equation}
where $P_V(V,T=0)$ is the zero temperature Vinet EOS \cite{vinet1987compressibility} given by
\begin{equation}
P_V = 3K_0x^{-2}(1-x)\exp [(1.5K'_0-1.5)(1-x)]
\end{equation}
where $x=(V/V_0)^{1/3}$ and 
\begin{equation}
P_{TH\text{ Debye}}(V,T) = \dfrac{9RT\gamma_D}{V}\left(\dfrac{T}{\theta_D}\right)^3 \int_0^{\theta_D/T} \dfrac{z^3}{e^z-1}dz.
\end{equation}
is the Debye thermal pressure. In total, the model has six parameters. The Vinet model has parameters $V_0$ the atomic volume, $K_0$ the bulk modulus, and $K'_0$ the pressure derivative of $K_0$ -- all at a reference state of ambient pressure and zero temperature. The Debye thermal pressure has an additional three parameters. The volume dependent characteristic Debye temperature $\theta_D$ is given by:
\begin{equation}
\theta_D = \theta_{D_0}x^{-1.5}\exp [\gamma_1(1-x^{3q})/q ]
\end{equation}
and the Debye-Gr{\"u}neisen parameter is given by 
\begin{equation}
\gamma_D = - \dfrac{d\ln \theta_D}{d \ln V} = \gamma_1x^{3q}+1/2.
\end{equation}
Fitting the model therefore requires a complicated calibration process to infer the following vector of six parameters $\alpha=\{ V_0, K_0, K'_0, \theta_0, \gamma_1, q\}$ from data at various pressures and temperatures \cite{dewaele2008high, occelli2003properties}. This calibration has been performed in the literature \cite{dewaele2008high} and we have further conducted a Bayesian parameter estimation using the same data with the resulting parameter distributions shown in Figure \ref{fig:MGD_Model}. These results were generated using Markov Chain Monte Carlo, MCMC, with an Affine-invariant sampler with Stretch moves \cite{goodman2010ensemble}, implemented in our UQpy software \cite{olivier2020uqpy, tsapetis2023uqpy}.

Our Bayesian parameter estimation, with naïve priors on the parameter values, produce well-converged posterior distributions and a high-quality fit to the experimental data. Yet the Bayesian calibration is allowed to accept values in regions where the parameters have no physical meaning; for example, the posterior distributions of the $\theta_{D_0}$ and $q$ parameters extend to negative values (Figure \ref{fig:MGD_Model}). These parameters describe the thermal contribution to the EOS, while data are limited to a small range of parameters in the $298^oK \le T \le 900^oK$ region, and so it can be expected that this model will quickly develop inconsistencies when extended over a large range of parameters. While this incorrect behavior would be relatively easy to correct through a prior that limits parameters to positive values, such an approach would be very difficult for more complex models where pathologies are difficult, if not impossible, to identify a priori. This is in addition to the well-known issues with scaling MCMC to large numbers of parameters and posterior distributions with complex structure as can be expected from poorly constrained problems.

The example in Figure \ref{fig:MGD_Model} serves to demonstrate a key difficulty with Bayesian model calibration, even when models have a modest number of parameters. Many physics-based and empirical EOS models have a much larger number of parameters which makes the parametric approach difficult and expensive.
For example, the Carbon EOS in the Radiative Emissivity and Opacity of Dense Plasmas (REODP) code \cite{miloshevsky2015atomic} has 17 parameters for the diamond phase of carbon (plus 17 each for the BC8, SC, SH phases and 30 parameters for the liquid phase), which makes even deterministic calibration a massive undertaking \cite{benedict2014multiphase}. Calibrating this set of parameters using Bayesian inference is practically impossible without huge data sets and highly specialized expertise.


\subsection{Thermodynamic constraints on EOS models}
\label{sec:constraints}
The EOS expresses the thermodynamic response of a material and so is subject to the laws of thermodynamics. These laws impose two types of constraints, often known as thermodynamic consistency and thermodynamic stability. The consistency constraint arises from the fact that changes in the various thermodynamic variables $\boldsymbol{\Theta}$ are all related to changes in a single quantity, the thermodynamic potential. For an EOS of the form in Eq.\ \eqref{eqn:EOS_Hyro}, this potential is the Helmholtz Free Energy given by:
\begin{equation}
    F = E - TS
\end{equation}
where $E$ is internal energy, $T$ is temperature, and $S$ is the entropy. According to the first and second laws of thermodynamics, changes in the state variables induce a change in the free energy $dF=-SdT-PdV$ allowing us to express the pressure and energy by:
\begin{equation}
\begin{aligned}
&P=-\left.\frac{\partial F}{\partial V}\right|_T,\\
&E=F-\left.T \frac{\partial F}{\partial T}\right|_V
\end{aligned}
\label{eqn:linopt}
\end{equation}
Taking the derivatives $\frac{\partial P}{\partial T}$ and $\frac{\partial E}{\partial V}$ gives the thermodynamic consistency constraint
\begin{equation}
P=T \frac{\partial P}{\partial T}-\frac{\partial E}{\partial V}.
\label{tc}
\end{equation}
Deviations from thermodynamic consistency represent an erroneous source or sink of heat or work in hydrocode simulations, and therefore any valid (useful) EOS model must satisfy this equality constraint. Thus, the space of functions that satisfy Eq.\ \eqref{tc} forms the maximal set of possible EOS functions for any system and therefore provides an upper bound on EOS uncertainty \cite{gaffney2022constraining}. Note that we have chosen the Helmholtz free energy to suit the data typically used to train EOS models (which have $T$ and $V$ as independent variables); the above discussion can be applied to any other choice of thermodynamic potential depending on the application.

The second EOS constraints, known as thermodynamic stability, are derived from the second law of thermodynamics that requires that the Helmholtz free energy is a convex function.  As a result, the isothermal compressibility ($\kappa_T$) and specific heat ($c_V$) are positive quantities, and the thermodynamic stability constraints are given as
\begin{eqnarray}
\left(\frac{\partial^2 F}{\partial V^2}\right)_T &&=-\left(\frac{\partial P}{\partial V}\right)_T=\frac{1}{V \kappa_T} \geq 0 \Longleftrightarrow \kappa_T>0 \Longleftrightarrow\left(\frac{\partial P}{\partial V}\right)_T \leqslant 0, 
\label{ts_P}
\end{eqnarray}
\begin{eqnarray}
\left(\frac{\partial^2 S}{\partial E^2}\right)_V&&=-\frac{1}{T^2}\left(\frac{\partial T}{\partial E}\right)_V=-\frac{1}{T^2 c_\nu} \leqslant 0 \Longleftrightarrow  c_V>0 \Longleftrightarrow\left(\frac{\partial E}{\partial T}\right)_V \geqslant 0 
\label{ts_E}
\end{eqnarray}
Deviations from thermodynamic stability can be catastrophic in hydrocode simulations, leading to numerical instability, and so the above convexity conditions provide another important constraint on the functional space of valid EOSs.

\subsection{Gaussian process regression}

The EOS model developed in this work is developed using physically constrained Gaussian process regression (GPR). GPR is a non-parametric supervised machine learning method that is widely used to construct surrogate models for expensive physics-based models \cite{giovanis2020data, fuhg2022local} since it can approximate complex non-linear functions with an inbuilt probabilistic estimate of prediction uncertainty. It is also easily interpretable, such that the Gaussian probability measure defined at each prediction point makes it straightforward to understand prediction uncertainty and establish a degree of confidence in the model. Furthermore, the model hyper-parameters establish the length-scale of the process and can be easily interpreted in terms of correlations among point predictions. Their flexibility allows them to cover a wide range of functional forms in a single model. These features of GPR makes it an ideal choice for quantifying model uncertainty. 

Formally, a GP is a stochastic process that is a collection of an infinite number of random variables indexed by time or space, such that any finite collection of these random variables $\left\{y=f(\mathbf{x}(\theta)) \mid \mathbf{x} (\theta) \in \mathbb{R}^d, \theta \in \Omega\right\}$ forms a multivariate Gaussian distribution. Hence, a GP can be completely defined by a joint Gaussian probability distribution over a set of functions \cite{williams2006gaussian}. A single function $f$ drawn from the set of admissible functions is known as a realization of the GP and, in our case, represents one possible model for the EOS. Our task in GPR is to identify the appropriate joint Gaussian probability distribution that best represents a set of available data.

Consider that, for a given set of $N$ observation of the input $\mathbf{x}$, i.e.\ $\mathbf{X}=\left\{\mathbf{x}^{(1)}, \mathbf{x}^{(2)}, \ldots \mathbf{x}^{(N)}\right\},\ \mathbf{x}^{(i)} \in \mathbb{R}^d$, we have the respective output vector $\mathbf{y}=\left[y^{(1)}, y^{(2)}, \ldots y^{(N)}\right]^{\top}, \ {y}^{(i)} \in \mathbb{R}$. 
We aim to use a GP to approximate the underlying function $Y(\cdot, \cdot): \mathbb{R}^d \times \Omega \rightarrow \mathbb{R}$.
In a Bayesian framework, we start by assuming a prior for GP $Y(\mathbf{x})$ as,
\begin{equation}
  Y(\mathbf{x}) \sim GP\left[\mu(\mathbf{x}), K\left(\mathbf{x}, \mathbf{x}^{\prime}\right)\right]  
\end{equation}
where $\mu(\cdot): \mathbb{R}^d \rightarrow \mathbb{R}$ and $K(\cdot, \cdot): \mathbb{R}^d \times \mathbb{R}^d \rightarrow \mathbb{R}$ are the mean function and covariance function, respectively, defined as,
\begin{eqnarray}
    &\mu(\mathbf{x})=\mathbb{E}[Y(\mathbf{x})]\\
    &    K\left(\mathbf{x}, \mathbf{x}^{\prime}\right)=\mathbb{E}[Y(\mathbf{x})-\mu(\mathbf{x})]\left[Y\left(\mathbf{x}^{\prime}\right)-\mu\left(\mathbf{x}^{\prime}\right)\right] \nonumber
\end{eqnarray}
The covariance function, selected as a positive definite kernel, defines the degree of linear dependence between the output values computed at input points $\mathbf{x}$ and $\mathbf{x}^{\prime}$. Typically, the closer two points are in the input space (by some measure, e.g.\ Euclidean distance), the more strongly correlated they are in the output space.  
A variety of kernel functions are available in the literature \cite{williams2006gaussian}. Throughout this work, we will use the square exponential covariance kernel with noise, given by
\begin{equation}
    K\left(\mathbf{x}, \mathbf{x}^{\prime}\right)=\sigma^2 \exp \left(\frac{-\left\|\mathbf{x}-\mathbf{x}^{\prime}\right\|_2^2}{2 l^2}\right)+\sigma_n^2 \delta_{\mathbf{x}, \mathbf{x}^{\prime}}
\label{kernel}
\end{equation}
where $l$, $\sigma^{2}$, $\sigma_{n}^{2}$, and $\delta_{x, x^{\prime}}$ are length-scale (correlation length, input scale), signal variance (output scale), Gaussian noise variance, and Kronecker delta function, respectively. Generally, $\theta = \left(\sigma, l, \sigma_{n} \right)$ denotes the set of hyper-parameters that are estimated from the training data. 

Next, define the matrix $\mathbf{K}=\left[K\left(\mathbf{x}^{(i)}, \mathbf{x}^{(j)}\right)\right]_{i j}$, the mean vector $  \boldsymbol{\mu}=\left[\mu\left(\mathbf{x}^{(1)}\right), \mu\left(\mathbf{x}^{(2)}\right), \ldots, \mu\left(\mathbf{x}^{(N)}\right)\right]^{\mathrm{T}}$, and the kernel entry $k\left(\mathbf{x}^{\prime}\right)=\left[K\left(\mathbf{x}^{(i)}, \mathbf{x}^{\prime}\right)\right]_i-\sigma_n^2 \delta_{\mathbf{x}^{(i)}, \mathbf{x}^{\prime}}$.
The posterior predictive distribution of the output $y^*$ for a new test input $\mathbf{x}^*$ conditioned on the training data set $(\mathbf{X},\mathbf{y})$ is given by,
\begin{equation}
  Y\left(\mathbf{x}^*\right) \mid \mathbf{y}, \mathbf{X} \sim \mathcal{N}\left[m\left(\mathbf{x}^*\right), s^2\left(\mathbf{x}^*\right)\right],   
\end{equation}
where
\begin{align}
    m\left(\mathbf{x}^*\right)&=\mu\left(\mathbf{x}^*\right)+k\left(\mathbf{x}^*\right)^{\mathrm{T}} \mathbf{K}^{-1}(\boldsymbol{y}-\boldsymbol{\mu}),\\
    s^2\left(\mathbf{x}^*\right)&= K\left(\mathbf{x}^*, \mathbf{x}^*\right)-k\left(\mathbf{x}^*\right)^{\mathrm{T}} \mathbf{K}^{-1} k\left(\mathbf{x}^*\right) .
\end{align}

The mean $m\left(\mathbf{x}^*\right)$ and variance $s^2\left(\mathbf{x}^*\right)$ are determined by estimating the hyper-parameters $\theta$. One popular approach to determine the optimal $\theta$ is to minimize the negative marginal log-likelihood \cite{williams2006gaussian}, given by 
\begin{equation}
-\log\ [p(\boldsymbol{y} \mid \boldsymbol{X}, \theta)]=\frac{1}{2}\left[(\boldsymbol{y}-\boldsymbol{\mu})^{\mathrm{T}} \mathbf{K}^{-1}(\boldsymbol{y}-\boldsymbol{\mu})+\log |\mathbf{K}|+N \ \log (2 \pi)\right]
\label{eqn:log-likelihood}
\end{equation}
This is performed by using a numerical optimizer.  For more details on standard GPR and its implementation, we refer the reader to the textbook by Williams and Rasmussen\cite{williams2006gaussian}. 

The standard GPR output is unconstrained, making it impractical for physics-based models such as EOS. Recently, several techniques have been developed to incorporate physical constraints on the GPR output \cite{swiler2020survey}. In our work, we employ two approaches to incorporate the thermodynamic consistency and stability constraints described in Section \ref{sec:constraints}. The first approach is based on the work by \citet{jidling2017linearly}, where they modify the kernel to incorporate the known linear operator constraints. We use this approach to design a specialized kernel that encodes the desired thermodynamic consistency constraint. The second approach
is recently proposed by \citet{pensoneault2020nonnegativity} to incorporate inequality-type constraints by minimizing the negative marginal log-likelihood function (Eq.\ \eqref{eqn:log-likelihood}) while requiring that the probability of violating the constraints is small. We use this approach to impose thermodynamic stability constraints. The mathematical details of incorporating these approaches in our proposed framework are described next.

\section{Mathematical formulation of the Constrained GP EOS}\label{Section 3}

In the following sections, we present a novel constrained GPR framework to build an uncertain EOS model constrained by the laws of thermodynamics. We first construct a GP EOS model constrained by thermodynamic consistency and stability constraints presented in Section \ref{sec:constraints}. We then use the resulting model to derive a GP model for the shock Hugoniot with uncertainty. Finally, we present a unified GPR that can be jointly trained from first-principles simulations and experimental shock Hugoniot data. 

\subsection{Thermodynamically constrained GP EOS model}\label{A}

Let $\mathbf{X}=\left\{\left(V,T\right)^{(i)}\right\}_{i=1}^N \in \mathcal{X}$ be $N$ input data points from the index set $\mathcal{X}$ with the corresponding EOS output $\mathbf{Y}= \left\{\left(P,E\right)^{(i)}\right\}_{i=1}^N$. We assume a GP prior for the Helmholtz free energy as
\begin{equation}
 F \sim GP\left[\mu_F(\mathbf{X}), k_{FF}\left(\mathbf{X}, \mathbf{X}^{\prime}\right)\right]   
 \label{eqn:F GP}
\end{equation}
Using Eq.\ \eqref{eqn:linopt}, we can define a linear operator as,
\begin{equation}
\mathcal{L}_{\mathbf{X}}=\left(\begin{array}{c}
-\frac{\partial}{\partial V} \\
1-T \frac{\partial}{\partial T}
\end{array}\right) 
\label{linearoperator}
\end{equation}
Since GPs are closed under linear operations, we can derive the joint GP priors for $P$ and $E$ as \cite{jidling2017linearly}, 
\begin{equation}
\left[\begin{array}{cc}
P \\
E
\end{array}\right] \mid \mathbf{X}\sim GP\left[\mathcal{L}_{\mathbf{X}}\mu_F(\mathbf{X}), \mathcal{L}_{\mathbf{X}}k_{FF}\left(\mathbf{X}, \mathbf{X}^{\prime}\right)\mathcal{L}_{\mathbf{X^{\prime}}}^{T}\right] 
\label{eqn:GP_PE}
\end{equation}
which can be rewritten as, 
\begin{equation}
\small
\left[\begin{array}{cc}
P \\
E
\end{array}\right] \mid \mathbf{X}\sim GP\left(\left[\begin{array}{cc}
\dfrac{\partial \mu_{F}\left(\mathbf{X}\right)}{\partial V} \\
\mu_{F}-T \dfrac{\partial \mu_{F}\left(\mathbf{X}\right)}{\partial T}
\end{array}\right],\left[\begin{array}{cc}
\dfrac{\partial}{\partial V}\dfrac{\partial}{\partial V^{\prime}}k_{FF}\left(\mathbf{X}, \mathbf{X}^{\prime}\right) & -\dfrac{\partial}{\partial V} \left(1-T^{\prime} \dfrac{\partial}{\partial T^{\prime}}\right) k_{FF}\left(\mathbf{X}, \mathbf{X}^{\prime}\right)\\
-\dfrac{\partial}{\partial V^{\prime}} \left(1-T \dfrac{\partial}{\partial T}\right) k_{FF}\left(\mathbf{X}, \mathbf{X}^{\prime}\right) & \left(1-T \dfrac{\partial}{\partial T}\right)\left(1-T^{\prime} \dfrac{\partial}{\partial T^{\prime}}\right) k_{FF}\left(\mathbf{X}, \mathbf{X}^{\prime}\right)
\end{array}\right]\right)
\label{eqn:joint_GP_PE}
\end{equation}
and ensures that the thermodynamic consistency constraint is guaranteed. For notational simplicity, let us denote Eq.\ \eqref{eqn:joint_GP_PE} as,
\begin{equation}
\left[\begin{array}{c}
P \\
E
\end{array}\right] \mid \mathbf{X}\sim GP\left(\left[\begin{array}{c}
\mu_{P}\left(\mathbf{X}\right) \\
\mu_{E}\left(\mathbf{X}\right)\\
\end{array}\right],\left[\begin{array}{cc}
K_{PP}\left(\mathbf{X}, \mathbf{X}\right) & K_{PE}\left(\mathbf{X}, \mathbf{X}\right)\\
K_{EP}\left(\mathbf{X}, \mathbf{X}\right) & K_{EE}\left(\mathbf{X}, \mathbf{X}\right)\\
\end{array}\right]\right)
\label{eqn:joint_GP}
\end{equation}
From this joint GP, the prediction
$P_{*}, E_{*}$ at a new point $\mathbf{X}_{*}$ can be calculated by conditioning as\\
\begin{equation}
\left[\begin{array}{c}
P \\
E\\
P_{*} \\
E_{*}
\end{array}\right] \mid \mathbf{X}, \mathbf{X}_{*}\sim GP\left(\left[\begin{array}{c}
\mu_{P}\left(\mathbf{X}\right) \\
\mu_{E}\left(\mathbf{X}\right)\\
\mu_{P}\left(\mathbf{X}_{*}\right) \\
\mu_{E}\left(\mathbf{X}_{*}\right)
\end{array}\right],\left[\begin{array}{cc|cc}
K_{PP}\left(\mathbf{X}, \mathbf{X}\right) & K_{PE}\left(\mathbf{X}, \mathbf{X}\right) &K_{PP}\left(\mathbf{X}, \mathbf{X}_{*}\right) & K_{PE}\left(\mathbf{X}, \mathbf{X}_{*}\right) \\
K_{EP}\left(\mathbf{X}, \mathbf{X}\right) & K_{EE}\left(\mathbf{X}, \mathbf{X}\right) & K_{EP}\left(\mathbf{X}, \mathbf{X}_{*}\right) & K_{EE}\left(\mathbf{X}, \mathbf{X}_{*}\right)\\
\hline K_{PP}\left(\mathbf{X}_{*}, \mathbf{X}\right) & K_{PE}\left(\mathbf{X}_{*}, \mathbf{X}\right) &K_{PP}\left(\mathbf{X}_{*}, \mathbf{X}_{*}\right) & K_{PE}\left(\mathbf{X}_{*}, \mathbf{X}_{*}\right) \\
K_{EP}\left(\mathbf{X}_{*}, \mathbf{X}\right) & K_{EE}\left(\mathbf{X}_{*}, \mathbf{X}\right) & K_{EP}\left(\mathbf{X}_{*}, \mathbf{X}_{*}\right) & K_{EE}\left(\mathbf{X}_{*}, \mathbf{X}_{*}\right)\\
\end{array}\right]\right)
\label{eqn:GP_prediction_PE}
\end{equation}
Again for notational simplicity, let us denote the block covariance matrix in Eq.\ \eqref{eqn:GP_prediction_PE} by
\begin{equation*}
    \left[\begin{array}{cc}
         \mathbf{K}_{11} & \mathbf{K}_{12}\\
         \mathbf{K}_{21} & \mathbf{K}_{22}
    \end{array} \right]
\end{equation*}
Further conditioning on the training data, we obtain the following GP model,
\begin{equation}
\begin{aligned}
\left[\begin{array}{c}
P_{*} \\
E_{*}
\end{array}\right] \mid {\mathbf{X}}, {\mathbf{X}}_*, P, E\sim GP\left(\left[\begin{array}{c}
\mu_{P}\left(\mathbf{X}_{*}\right) \\
\mu_{E}\left(\mathbf{X}_{*}\right)
\end{array}\right]+\mathbf{K}_{21}\left(\mathbf{K}_{11}\right)^{-1}\bigg(\left[\begin{array}{c}
P \\
E
\end{array}\right]-\left[\begin{array}{c}
\mu_{P}\left(\mathbf{X}\right) \\
\mu_{E}\left(\mathbf{X}\right)
\end{array}\right]\bigg),\ \mathbf{K}_{22}-\mathbf{K}_{21}\left(\mathbf{K}_{11}\right)^{-1} \mathbf{K}_{12}\right)
\end{aligned}
\label{eqns:PE_pred}
\end{equation}\\
The negative log marginal likelihood of the joint GP is given by, 
\begin{multline}
-\log p\left(P, E \mid \mathbf{X}, \boldsymbol{\theta}\right)=\frac{1}{2}\left(\left[\begin{array}{c}
P \\
E
\end{array}\right]-\left[\begin{array}{c}
\mu_{P}\left(\mathbf{X}\right) \\
\mu_{E}\left(\mathbf{X}\right)
\end{array}\right]\right)^{T} \mathbf{K}_{\text {11}}^{-1}\left(\left[\begin{array}{c}
P \\
E
\end{array}\right]-\left[\begin{array}{c}
\mu_{P} \\
\mu_{E}
\end{array}\right]\right)+\\ 
\frac{1}{2} \log \left|\mathbf{K}_{\text {11}}\right|+\frac{N}{2} \log (2 \pi)
\label{eqn: NLML}
\end{multline}

Next, we enforce the thermodynamic stability constraints (Eqs.\ \eqref{ts_P} and \eqref{ts_E}) by limiting the functional space through constrained hyper-parameter optimization \cite{pensoneault2020nonnegativity}. We obtain the hyper-parameters by minimizing the negative marginal log-likelihood function in Eq.\ \eqref{eqn: NLML} while requiring that the probability of violating the thermodynamics stability constraints is small. Formally, for $0 <\eta \ll 1$, we impose the following probabilistic constraints at virtual locations in the input domain $\mathbf{X}_v$ as, 
\begin{equation}
P\left[\left(\dfrac{\partial P\left(\mathbf{X}_v\right)}{\partial V} \mid \mathbf{X}_v, P, \mathbf{X}\right)>0\right] \leq \eta, \quad \text { for all } \mathbf{X}_v \in \mathcal{X}
\label{prob_const1}
\end{equation}
\begin{equation}
P\left[\left(\dfrac{\partial E\left(\mathbf{X}_v\right)}{\partial T} \mid \mathbf{X}_v, E, \mathbf{X}\right)<0\right] \leq \eta, \quad \text { for all } \mathbf{X}_v \in \mathcal{X}
\label{prob_const2}
\end{equation}
Since $\frac{\partial P\left(\mathbf{X}_v\right)}{\partial V} \mid \mathbf{X}_v, P, \mathbf{X}$ and $\frac{\partial E\left(\mathbf{X}_v\right)}{\partial T} \mid \mathbf{X}_v, E, \mathbf{X}$ follow a Gaussian distribution, the constraints in Eq.\ \eqref{prob_const1} and \eqref{prob_const2}, can be simplified as, 
\begin{equation}
    \mu_{\frac{\partial P}{\partial V}} - \Phi^{-1}(\eta) \ \sigma_{\frac{\partial P}{\partial V}} \leq 0
\label{stability1}
\end{equation}
and
\begin{equation}
    \mu_{\frac{\partial E}{\partial T}} +\Phi^{-1}(\eta) \ \sigma_{\frac{\partial E}{\partial T}} \geq 0
\label{stability2}
\end{equation}
where $\Phi^{-1}$ is the inverse standard normal cumulative distribution function. By minimizing the objective function Eq.\ \eqref{eqn: NLML} subject to constraints Eq.\ \eqref{stability1} and Eq.\ \eqref{stability2}, we can obtain a set of hyper-parameters, $\theta$, that ensures the resulting GP EOS model (Eq.\ \eqref{eqns:PE_pred}) satisfies both the thermodynamic consistency and stability constraints.

\subsection{Hugoniot derivation from the GP EOS model}\label{B}

In this section, we first derive the Hugoniot function ($H$) as a GP from the constrained GP EOS model described in Section \ref{A}. Then, we obtain the probabilistic set of so-called Hugoniot points satisfying $H (V, T)=0$. 


From Eqs.\ \eqref{eqns:meanH} and \eqref{eqns:covHH}, the $H$ GP prior is given as 
\begin{equation}
 H \sim GP\left[\mu_H(\mathbf{X}), K_{HH}\left(\mathbf{X}, \mathbf{X}^{\prime}\right)\right]   
 \label{eqn:H_GP}
\end{equation}
The predictive distribution of $H$ at test points $\mathbf{X}_{*}$ is given by
\begin{multline}
H_{*} \mid {\mathbf{X}}, {\mathbf{X}}_*, H \sim GP  (
\mu_{H}\left(\mathbf{X}_{*}\right) + K_{HH}\left(\mathbf{X}_{*}, \mathbf{X}\right)(K_{HH}\left(\mathbf{X}, \mathbf{X}\right))^{-1}
(H - \mu_{H}\left(\mathbf{X}\right)),\\
 K_{HH}\left(\mathbf{X}_{*}, \mathbf{X}_{*}\right)-K_{HH}\left(\mathbf{X}_{*}, \mathbf{X}\right)(K_{HH}\left(\mathbf{X}, \mathbf{X}\right))^{-1} K_{HH}\left(\mathbf{X}, \mathbf{X}_{*}\right)),
\label{eqns:H_pred}
\end{multline}
Using Eq.\ \eqref{eqns:H_pred}, we can define a subset $\mathcal{X}_H \subset \mathcal{X}$ such that $\forall$ $\mathbf{X}_H \in \mathcal{X}_H$, $H(\mathbf{X}_H)=0$ lies within the $1-\alpha$\% confidence intervals of the GP for $H$. We can achieve this by defining the standardized GP
\begin{equation}
\widetilde{H}(\mathbf{X}) = \dfrac{H(\mathbf{X})-\mu_H(\mathbf{X})}{\sigma_H(\mathbf{X})}
\end{equation}
where $\mathbf{X}_H$ satisfies $P(|\widetilde{H}(\mathbf{X}_H)|\le z_{\alpha/2})=1-\alpha$.
In other words, the points $\mathbf{X}_H$ satisfy the following condition
\begin{equation}
    |\mu_H(\mathbf{X}_H)|\leq z_{\alpha/2} \times \sigma_H(\mathbf{X}_H).
\label{hugo}
\end{equation}
Given an arbitrary point $X_H\in\mathcal{X}$, we can therefore establish the predictive distribution for pressure and internal energy at this point, $P_H$ and $E_H$ using Eq.\ \eqref{eqns:PE_pred}; thus providing an estimate of the uncertain Hugoniot curve satisfying $H(X_H)=0$.  

\subsection{Unified GP EOS model learned from multiple data sources}\label{C}
In this section, we propose a unified framework to train the proposed constrained GP EOS model using heterogeneous data sources. In particular, we show that the EOS model can be learned from a combination of first-principles simulation data and experimental shock Hugoniot observations. Let us define the model outputs as $P$, $E$, and $H$, for respective inputs, $\mathbf{X}_P$, $\mathbf{X}_E$, and $\mathbf{X}_H$.
The joint GP of $P, E, H$ is then defined as follows\\
\begin{equation}
\left[\begin{array}{c}
P \\
E\\
H
\end{array}\right] \mid \mathbf{X}_{P}, \mathbf{X}_{E}, \mathbf{X}_{H}\sim GP\left(\left[\begin{array}{c}
\mu_{P}\left(\mathbf{X}_{P}\right) \\
\mu_{E}\left(\mathbf{X}_{E}\right)\\
\mu_{H}\left(\mathbf{X}_{H}\right)
\end{array}\right],\left[\begin{array}{ccc}
K_{PP}\left(\mathbf{X}_{P}, \mathbf{X}_{P}\right) & K_{PE}\left(\mathbf{X}_{P}, \mathbf{X}_{E}\right) & K_{PH}\left(\mathbf{X}_{P}, \mathbf{X}_{H}\right) \\
K_{EP}\left(\mathbf{X}_{E}, \mathbf{X}_{P}\right) & K_{EE}\left(\mathbf{X}_{E}, \mathbf{X}_{E}\right) & K_{EH}\left(\mathbf{X}_{E}, \mathbf{X}_{H}\right) \\
K_{HP}\left(\mathbf{X}_{H}, \mathbf{X}_{P}\right) & K_{HE}\left(\mathbf{X}_{H}, \mathbf{X}_{E}\right) & K_{HH}\left(\mathbf{X}_{H}, \mathbf{X}_{H}\right)\\
\end{array}\right]\right)
\label{eqn:joint_GP}
\end{equation}
We can train the joint GP using any combination of available data and impose the thermodynamic constraints similar to the steps described in Section \ref{A} to obtain the predictive distribution of the joint GP of $P$, $E$, and $H$ at any test point $\mathbf{X}_{*}$.
Further, we can condition on $H$ by first partitioning the block covariance of the joint GP in Eq.\ \eqref{eqn:joint_GP} in a similar manner as done previously.
We denote this block covariance by
\begin{equation*}
    \left[\begin{array}{cc}
         \mathbf{K}_{EP} & \mathbf{K}_{*H}\\
         \mathbf{K}_{H*} & \mathbf{K}_{HH}
    \end{array} \right].
\end{equation*}
Conditioning on $H$ gives the resulting conditional distribution
\begin{multline}
\left[\begin{array}{c}
P \\
E\\
\end{array}\right] \mid \mathbf{X}_{P}, \mathbf{X}_{E}, \mathbf{X}_{H}, H(\mathbf{X}_{H})=0 \sim GP\bigg(\textbf{m}(\mathbf{X})=\left[\begin{array}{c}
\mu_{P}\left(\mathbf{X}_{P}\right) \\
\mu_{E}\left(\mathbf{X}_{H}\right)\\
\end{array}\right]+\mathbf{K}_{*H}\left(\mathbf{K}_{HH}\right)^{-1}\Big(H(\mathbf{X}_{H}) - \mu_{H}\left(\mathbf{X}_{H}\right)\Big),\ \\
\boldsymbol{\Sigma}_{\mathbf{X}\mathbf{X}}= \mathbf{K}_{EP}-\mathbf{K}_{*H}\left(\mathbf{K}_{HH}\right)^{-1} \mathbf{K}_{H*}\bigg)
\end{multline}

Next, conditioning on a set of training points $\mathbf{X}_H$ constrained by $H(\mathbf{X}_H)=0$ (e.g. from experimental data collected along the Hugoniot curve) yields
\begin{equation}
\left[\begin{array}{c}
P \\
E\\
P_H \\
E_H\\
\end{array}\right] \mid \mathbf{X}_{P}, \mathbf{X}_{E}, \mathbf{X}_{H}, H(\mathbf{X}_{H})=0 \sim GP\left(\left[\begin{array}{c}
\textbf{m}\left(\mathbf{X}\right) \\
\textbf{m}\left(\mathbf{X}_H\right) \\
\end{array}\right],\left[\begin{array}{cc}
\boldsymbol{\Sigma}_{\mathbf{X}\mathbf{X}} &  \boldsymbol{\Sigma}_{\mathbf{X}\mathbf{X}_{H}}\\
\boldsymbol{\Sigma}_{\mathbf{X}_{H}\mathbf{X}}  & \boldsymbol{\Sigma}_{\mathbf{X}_{H}\mathbf{X}_{H}} \\
\end{array}\right]\right)
\label{eqns:condH}
\end{equation}
Eq.\ \eqref{eqns:condH} yields the predictive distribution of $P$ and $E$ at $\mathbf{X}_{H}$ given by
\begin{multline}
\left[\begin{array}{c}
P_H \\
E_H\\
\end{array}\right] \mid \mathbf{X}_{P}, \mathbf{X}_{E}, \mathbf{X}_{H},P, E, H(\mathbf{X}_{H})=0 \sim  GP\bigg(\textbf{m}(\mathbf{X}_{H})+\boldsymbol{\Sigma}_{\mathbf{X}_{H}\mathbf{X}}\left(\boldsymbol{\Sigma}_{\mathbf{X}\mathbf{X}}\right)^{-1}\Big(\left[\begin{array}{c}
P \\
E\\
\end{array}\right]-\left[\begin{array}{c}
\mu_{P}\left(\mathbf{X}_{P}\right) \\
\mu_{E}\left(\mathbf{X}_{E}\right)\\
\end{array}\right]\Big),\ \\ \boldsymbol{\Sigma}_{\mathbf{X}_{H}\mathbf{X}_{H}}-\boldsymbol{\Sigma}_{\mathbf{X}_{H}\mathbf{X}}\left(\boldsymbol{\Sigma}_{\mathbf{X}\mathbf{X}}\right)^{-1} \boldsymbol{\Sigma}_{\mathbf{X}\mathbf{X}_{H}}\bigg).
\end{multline}
Using these equations, it is now possible to learn the thermodynamically constrained GP EOS model using a combination of experimentally observed points along the shock Hugoniot and first-principles calculations that relate $P,V,T,E$ as we demonstrate in the next section.

\section{Results and discussion}
In this section, we apply the proposed constrained GPR framework described in Section \ref{Section 3} to learn the EOS for the diamond phase of Carbon. We use a sample of 20 data points obtained from Density Functional Theory Molecular Dynamics (DFT-MD) simulations from Benedict et al.\ \cite{benedict2014multiphase} to train the GP EOS model. We first build an uncertain EOS model that satisfies both the thermodynamic consistency and stability constraints. We then derive the Hugoniot function GP from the EOS model and obtain the resulting shock Hugoniots with uncertainty. Finally, we train the unified GP EOS model using the simulation and a limited number of experimental Hugoniot data from laser-driven shock compression experiments \cite{mcwilliams2010strength}.

\subsection{Constrained GP EOS model for Diamond}

The EOS model is trained by assuming a squared exponential covariance model (Eq.\ \eqref{kernel}) for the Helmholtz free energy and computing the mean and covariance function matrix of the joint GP for pressure and energy $(P,E)$ from Eq.\ \eqref{eqn:joint_GP_PE} with input states of volume and temperature $(V,T)$.
This procedure ensures that the thermodynamic consistency constraint is satisfied. We then perform constrained hyper-parameter optimization using the COBYLA optimizer from the SCIPY python package \cite{virtanen2020scipy} to incorporate the probabilistic thermodynamic stability constraints from Eqs.\ \eqref{stability1} and \eqref{stability2} with $\eta = 0.025$. 

\begin{figure}[!ht]
   \centering 
    \begin{tabular}{cc}
\includegraphics[width=0.45\textwidth]{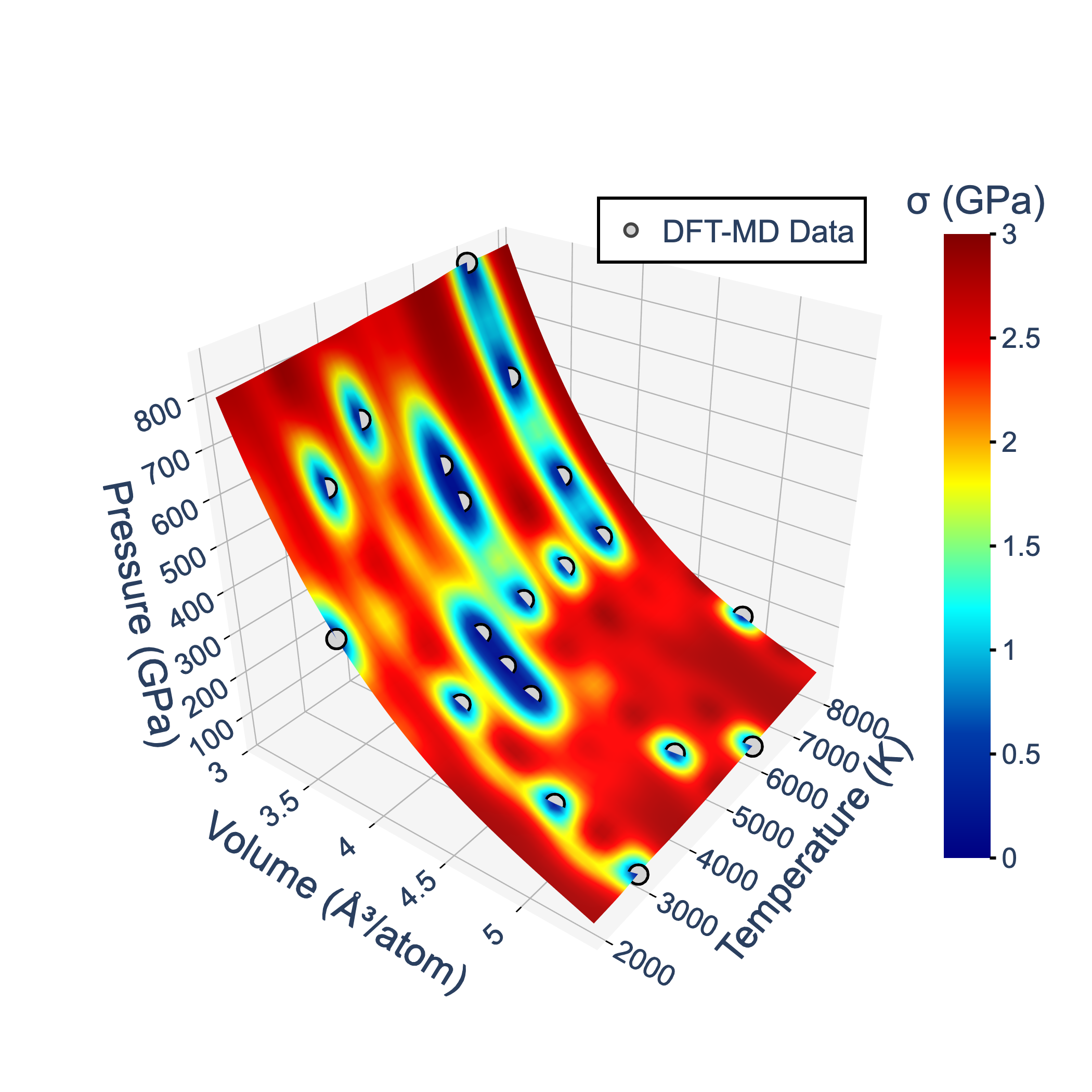} &
\includegraphics[width=0.45\textwidth]{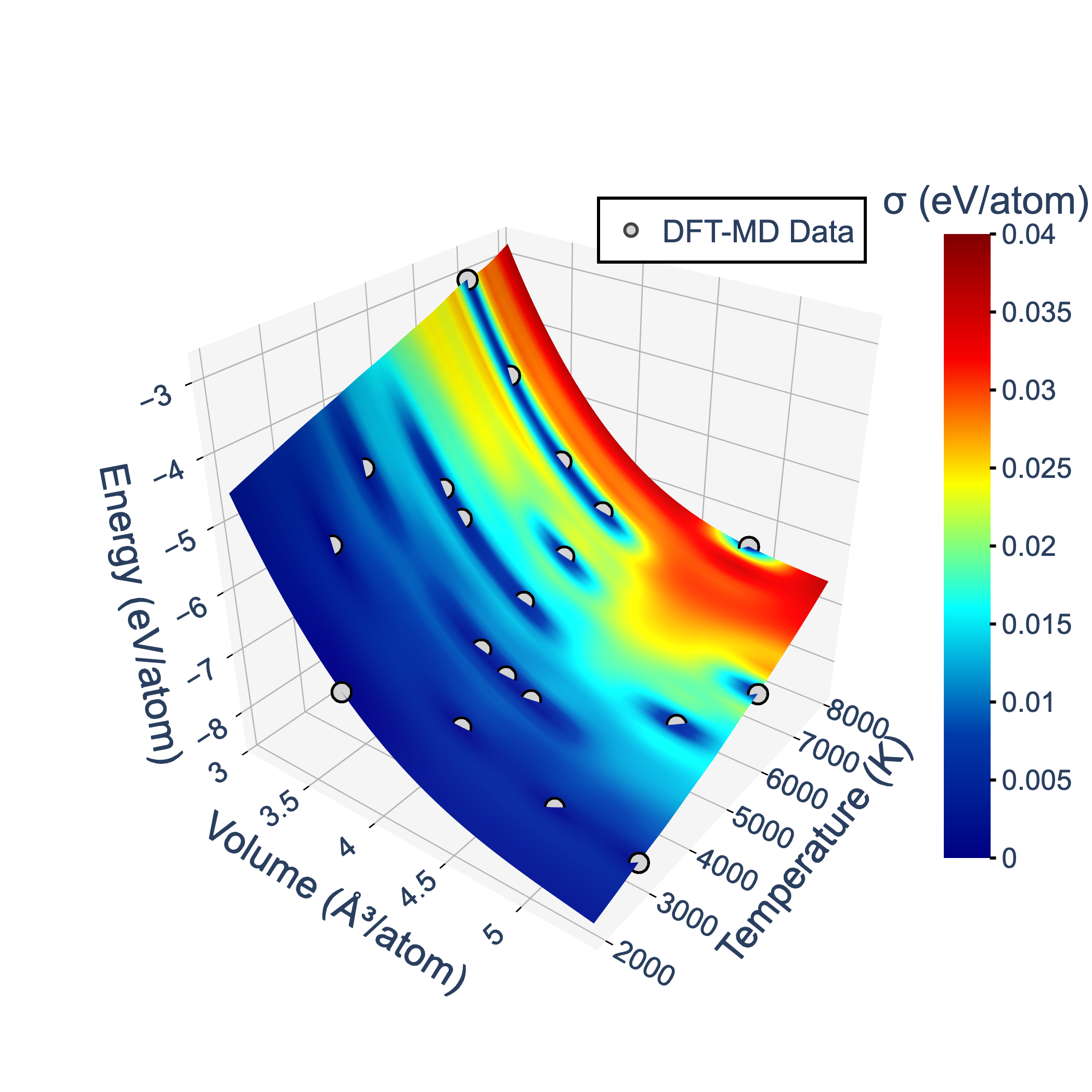} \\ 
(a)& (b)\\
\includegraphics[width=0.45\textwidth]{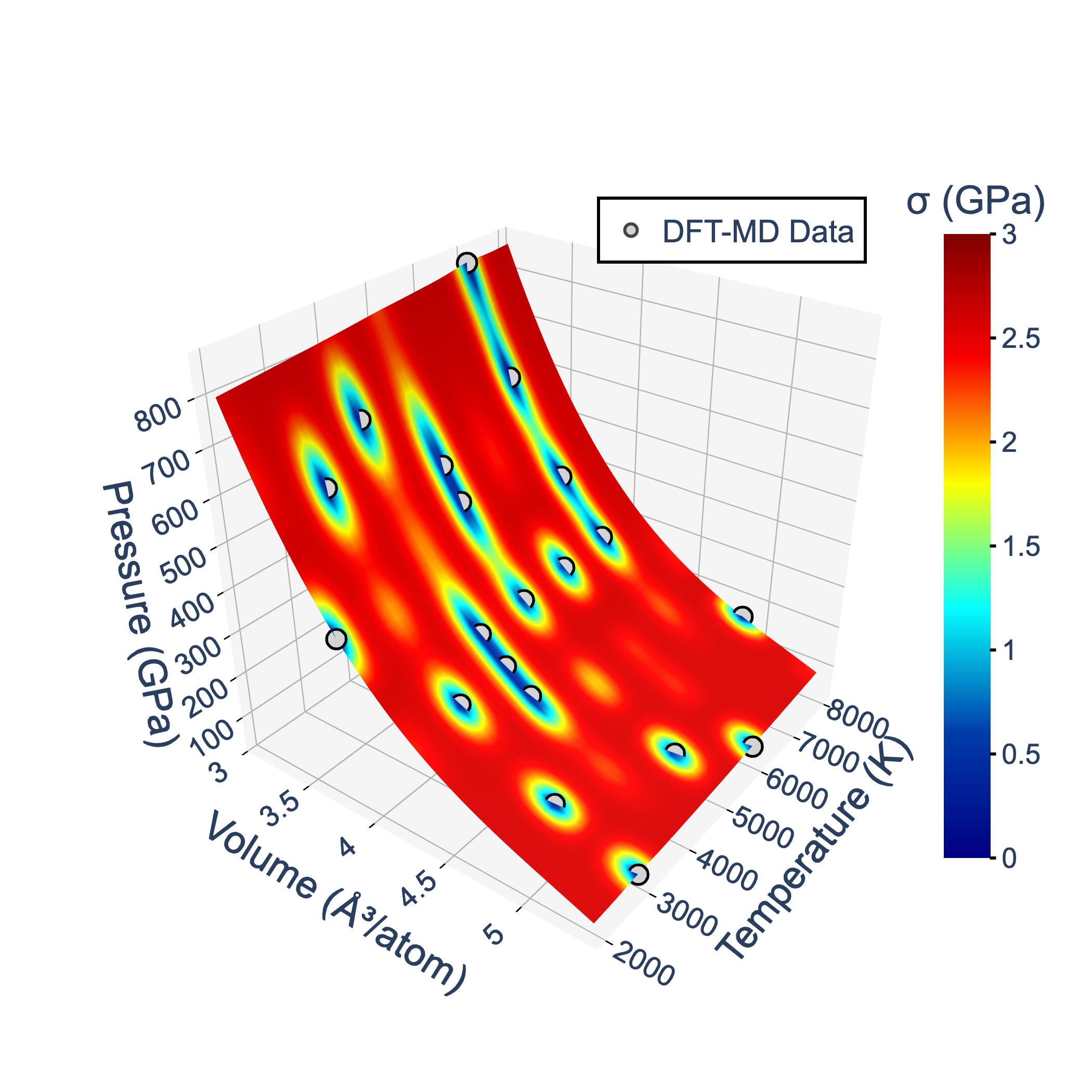} &
\includegraphics[width=0.45\textwidth]{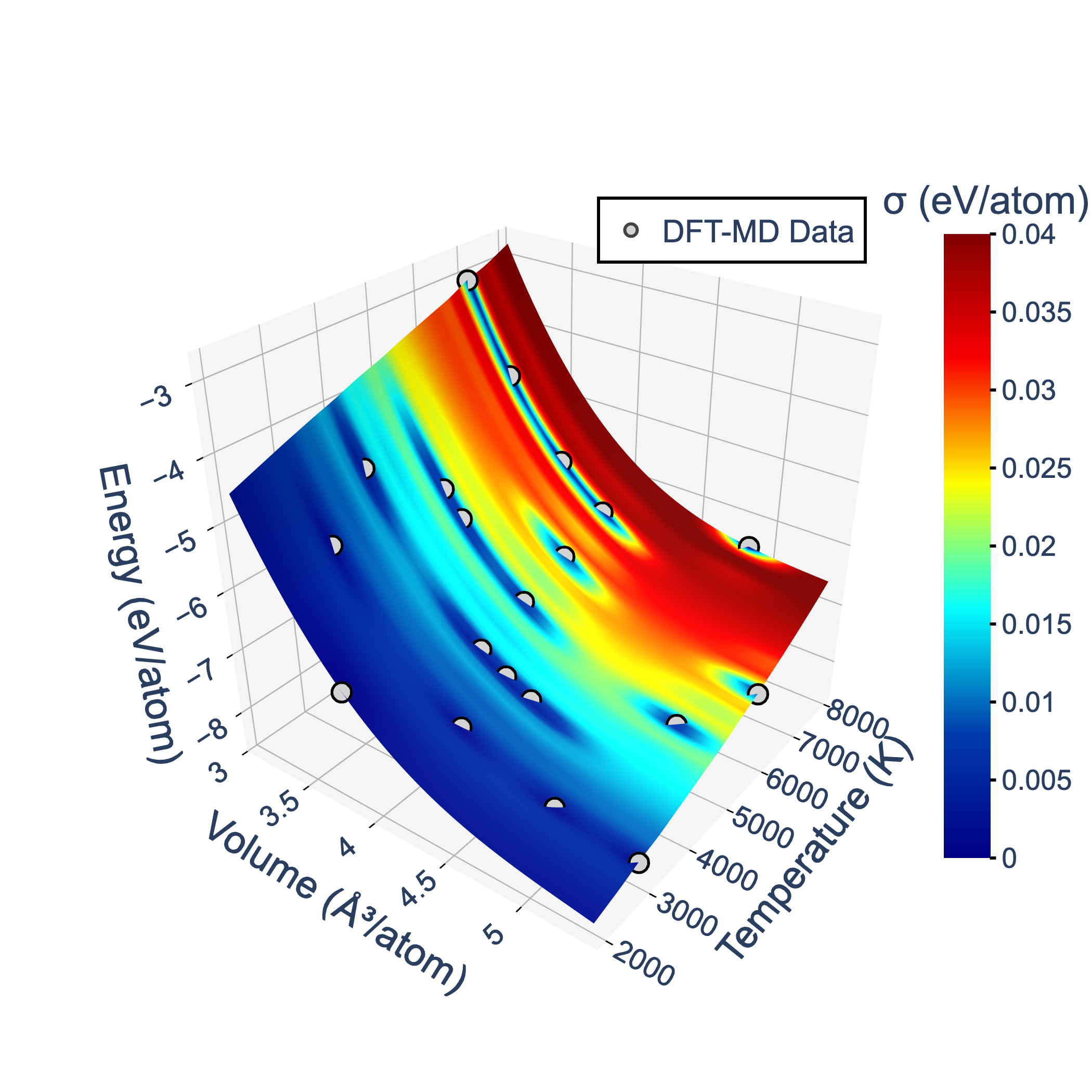} \\ 
(c)& (d)\\
\end{tabular}
\caption{Joint GP EOS model with uncertainty with and without thermodynamic constraints. (a) Constrained pressure marginal GP model. (b) Constrained energy marginal GP model. (c) Unconstrained pressure marginal GP model. (d) Unconstrained energy marginal GP model. Colors denote the standard deviation of the model at each point.}
\label{fig:constrainedPE-GP}
\end{figure}

Figures \ref{fig:constrainedPE-GP}(a) and (b) show the marginalized pressure GP (P-GP) and energy GP (E-GP) as a function of state variables $(V, T)$, respectively, trained from 20 DFT-MD simulations where uncertainties are shown with colors denoting standard deviation. We clearly see that the uncertainties are small near the training data points and larger in regions with no data. The overall uncertainty is small, with the coefficient of variation (COV) under $7\%$ for the P-GP and under $1.3\%$ for E-GP, suggesting high confidence in the predicted state. 
We similarly trained a GP model without imposing constraints 
as shown in Figure \ref{fig:constrainedPE-GP}(c) and (d), which show that the unconstrained EOS model has higher uncertainty in general. 

\begin{figure}[t!]
   \centering 
    \begin{tabular}{cc}
\includegraphics[width=0.45\textwidth]{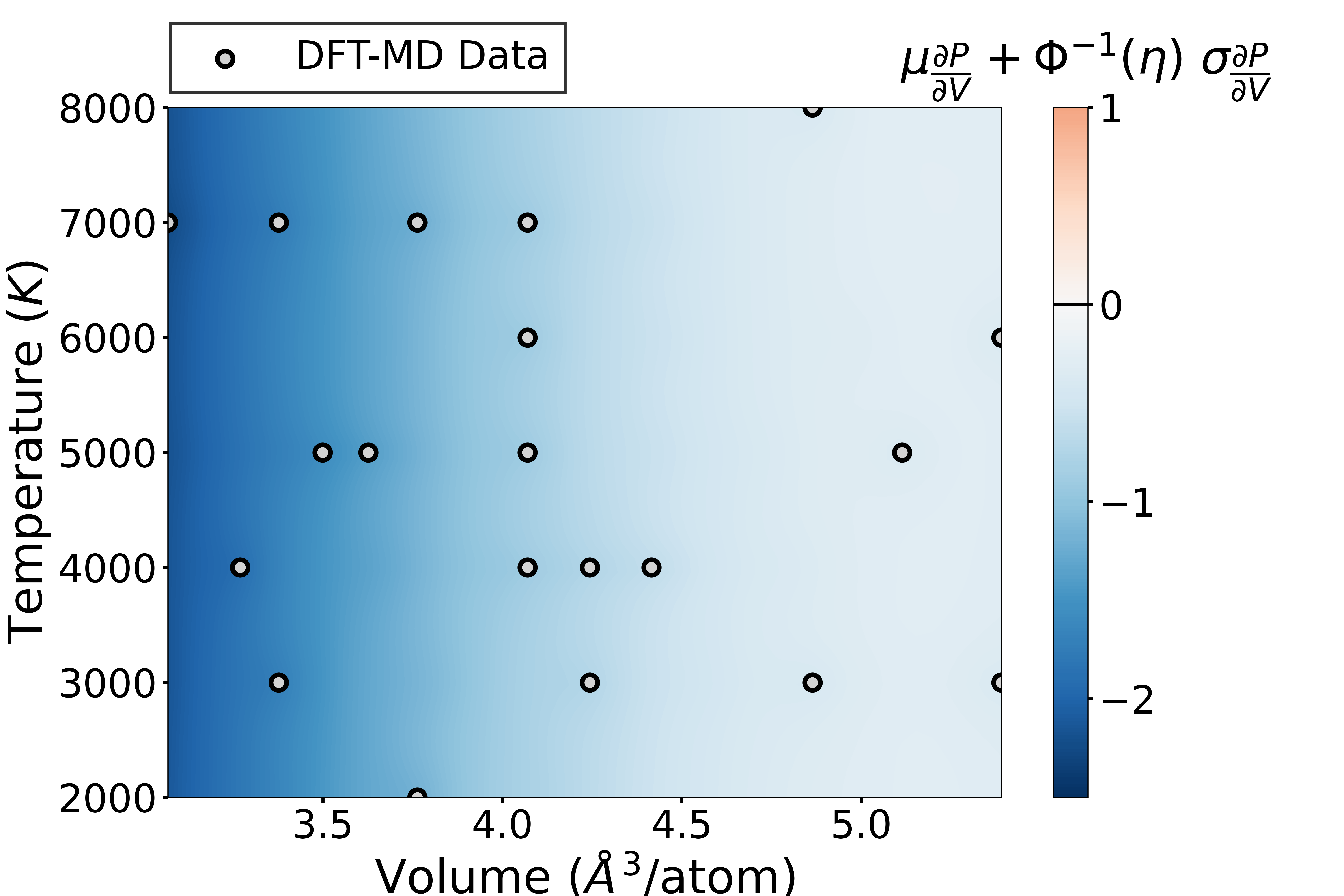} &
\includegraphics[width=0.45\textwidth]{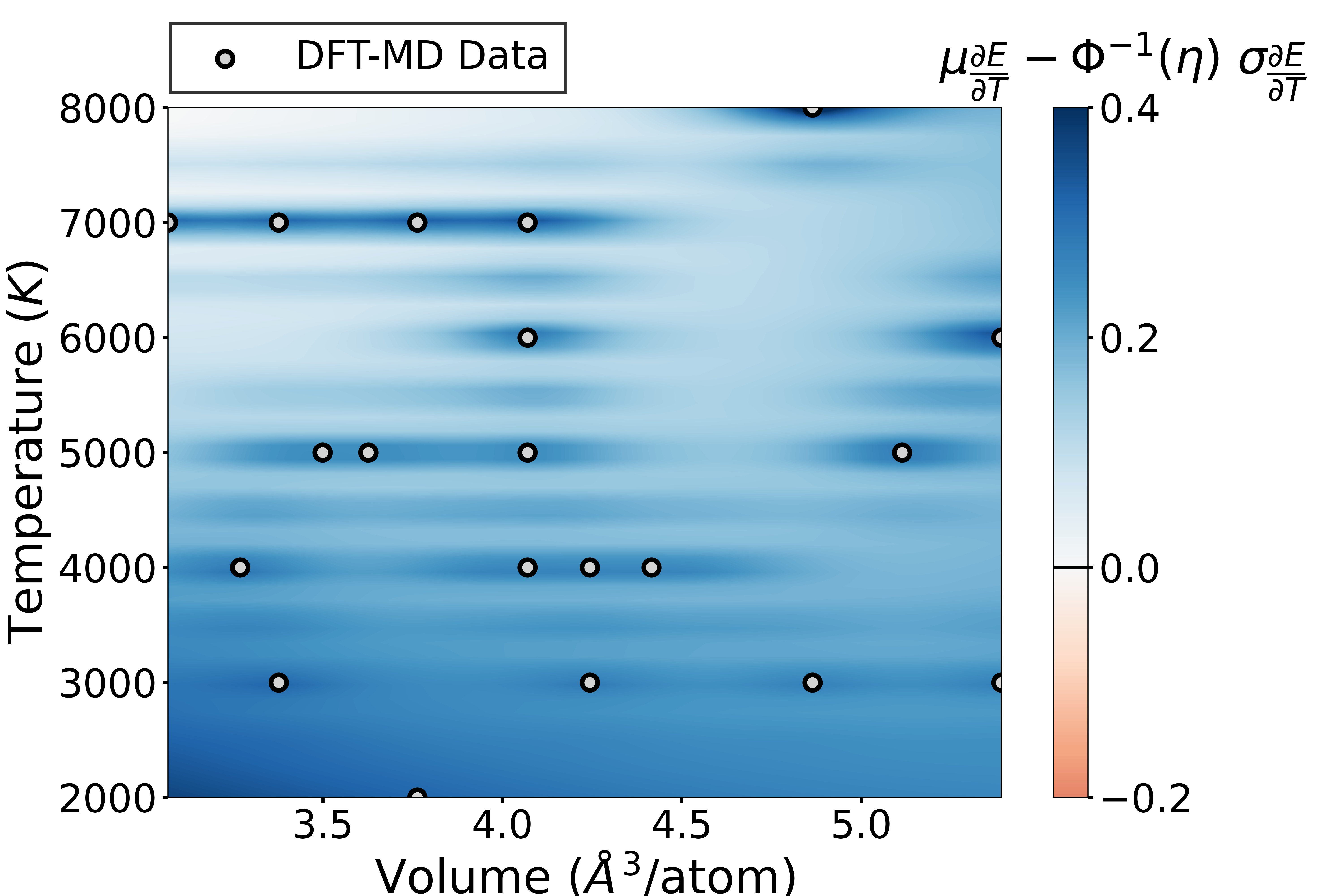} \\ 
(a)& (b)\\
\includegraphics[width=0.45\textwidth]{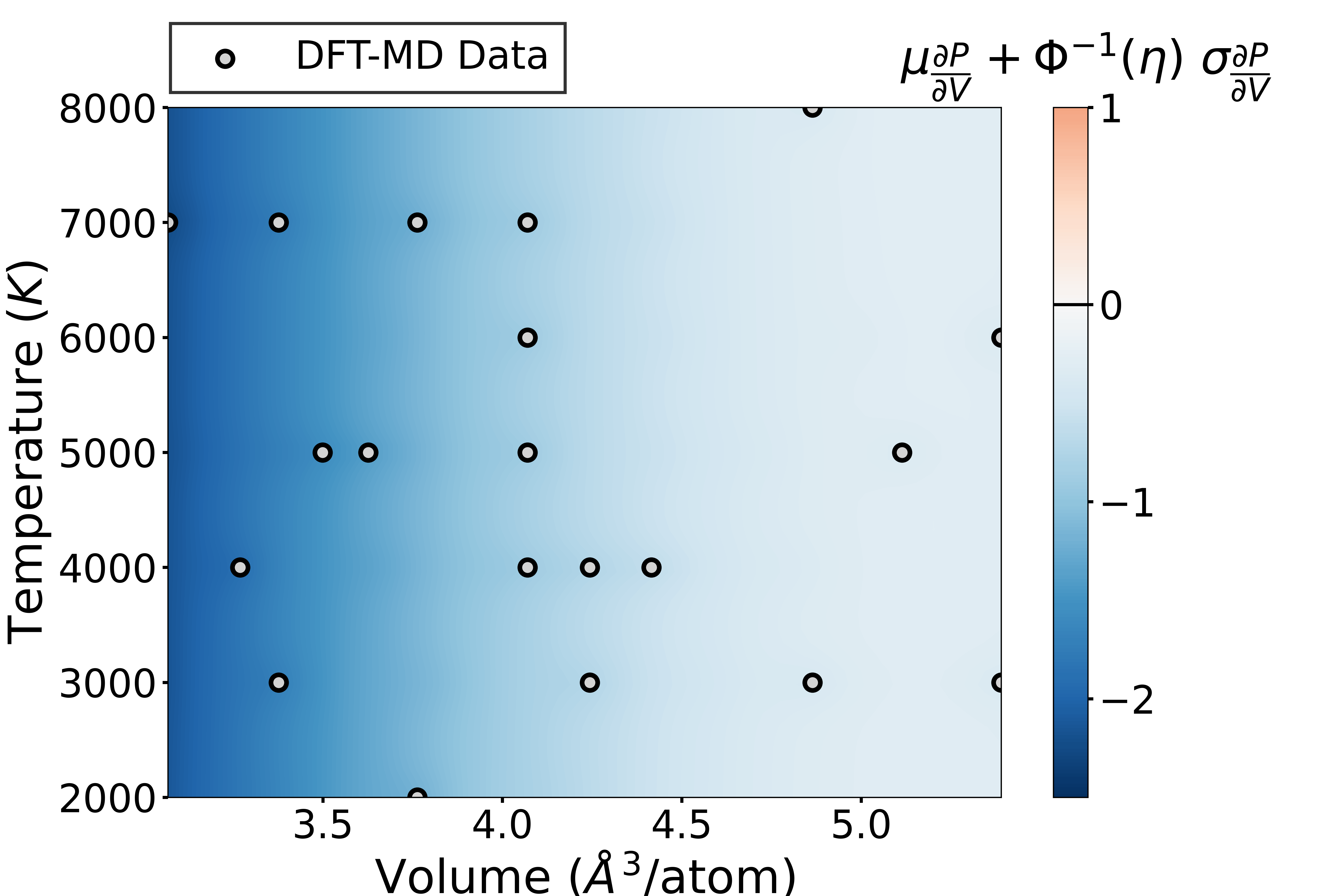} &
\includegraphics[width=0.45\textwidth]{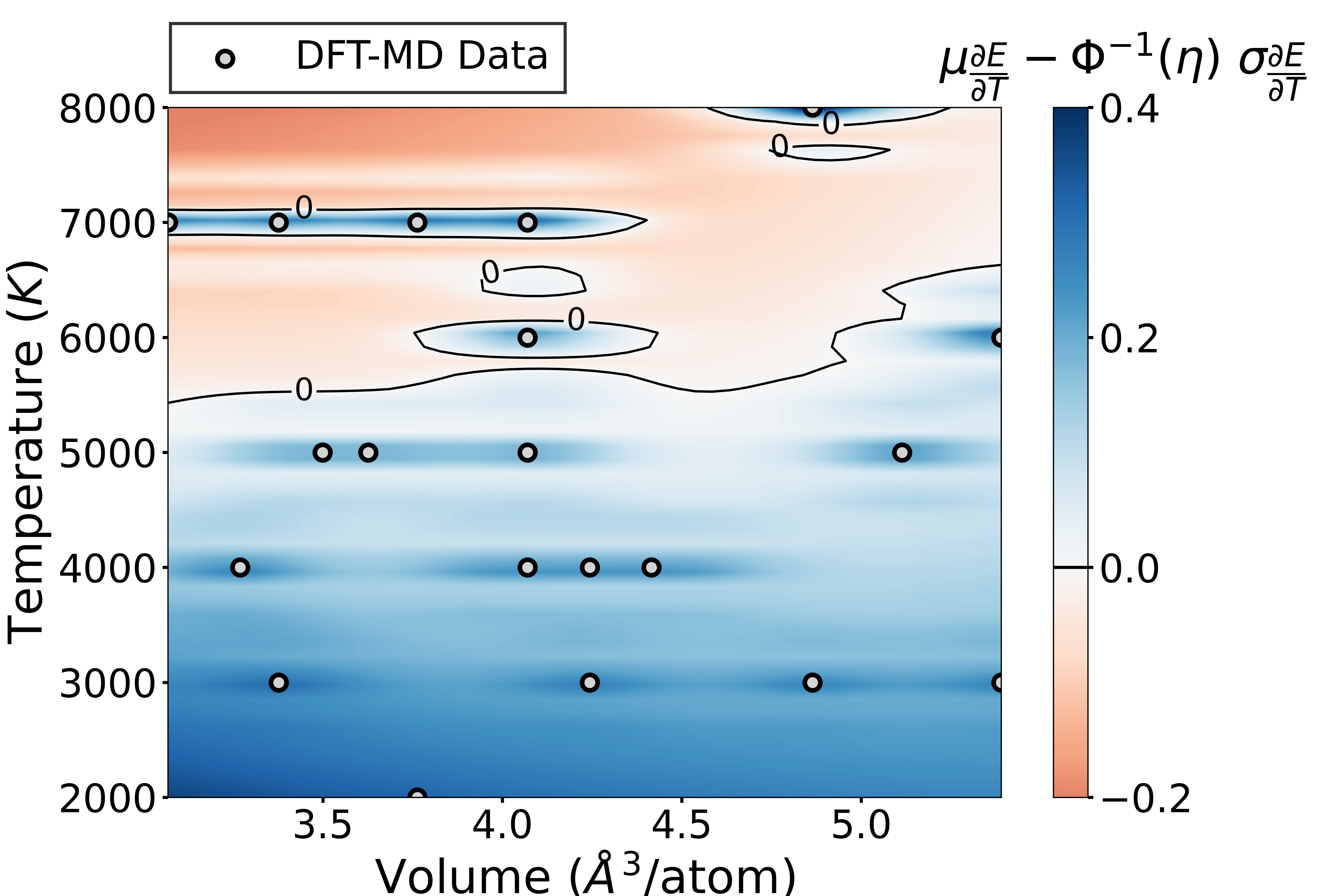} \\ 
(c)& (d)\\
\end{tabular}
\caption{Plots of the thermodynamic stability conditions in the $(V,T)$ space. Regions of constraint violation are represented by red. (a) Probabilistic stability constraint for constrained Pressure GP EOS model showing that the constraints are not violated. (b) Probabilistic stability constraint for constrained Energy GP EOS model showing that the constraints are not violated. (c) Probabilistic stability constraint for unconstrained Pressure GP EOS model showing that this specific constraint is not violated even when unconstrained. (d) Probabilistic stability constraint for unconstrained Energy GP EOS model showing that the constraint is violated in regions characterized by high temperatures across volumes within the domain.}
\label{fig:constraint_plots}
\end{figure}
\begin{figure}[t!]
   \centering 
    \begin{tabular}{cc}
\includegraphics[width=0.45\textwidth]{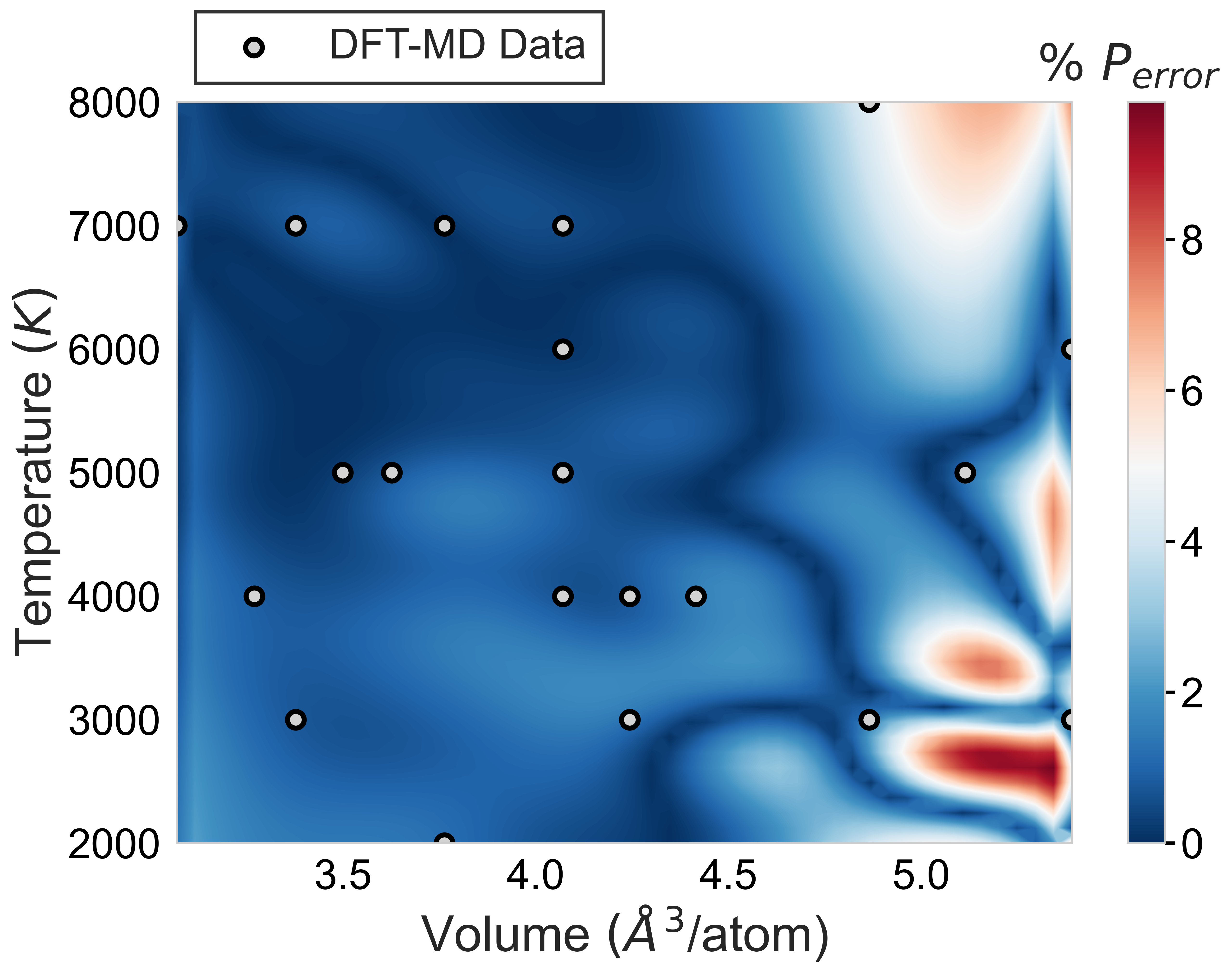} &
\includegraphics[width=0.45\textwidth]{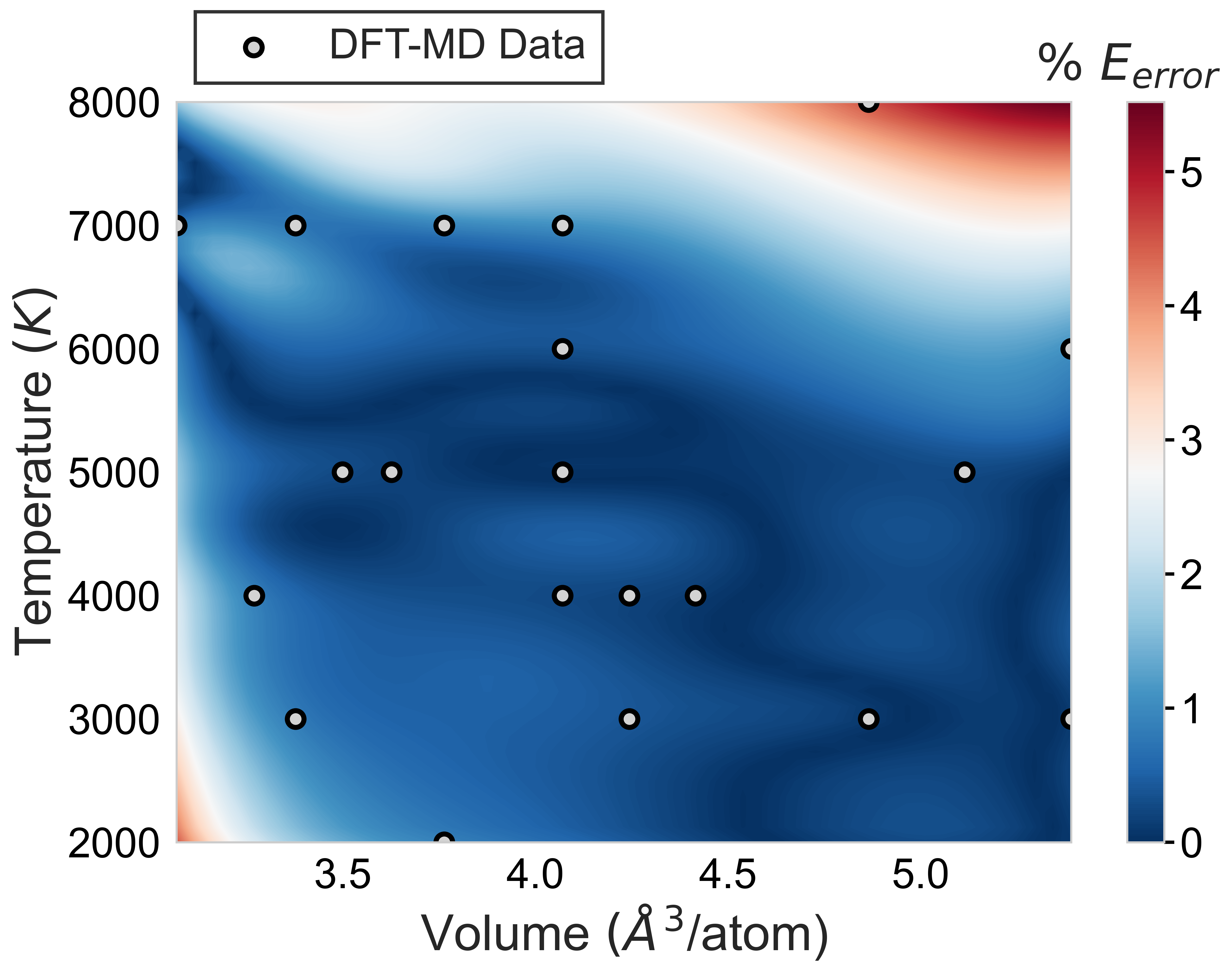} \\ 
(a)& (b)\\
\includegraphics[width=0.48\textwidth]{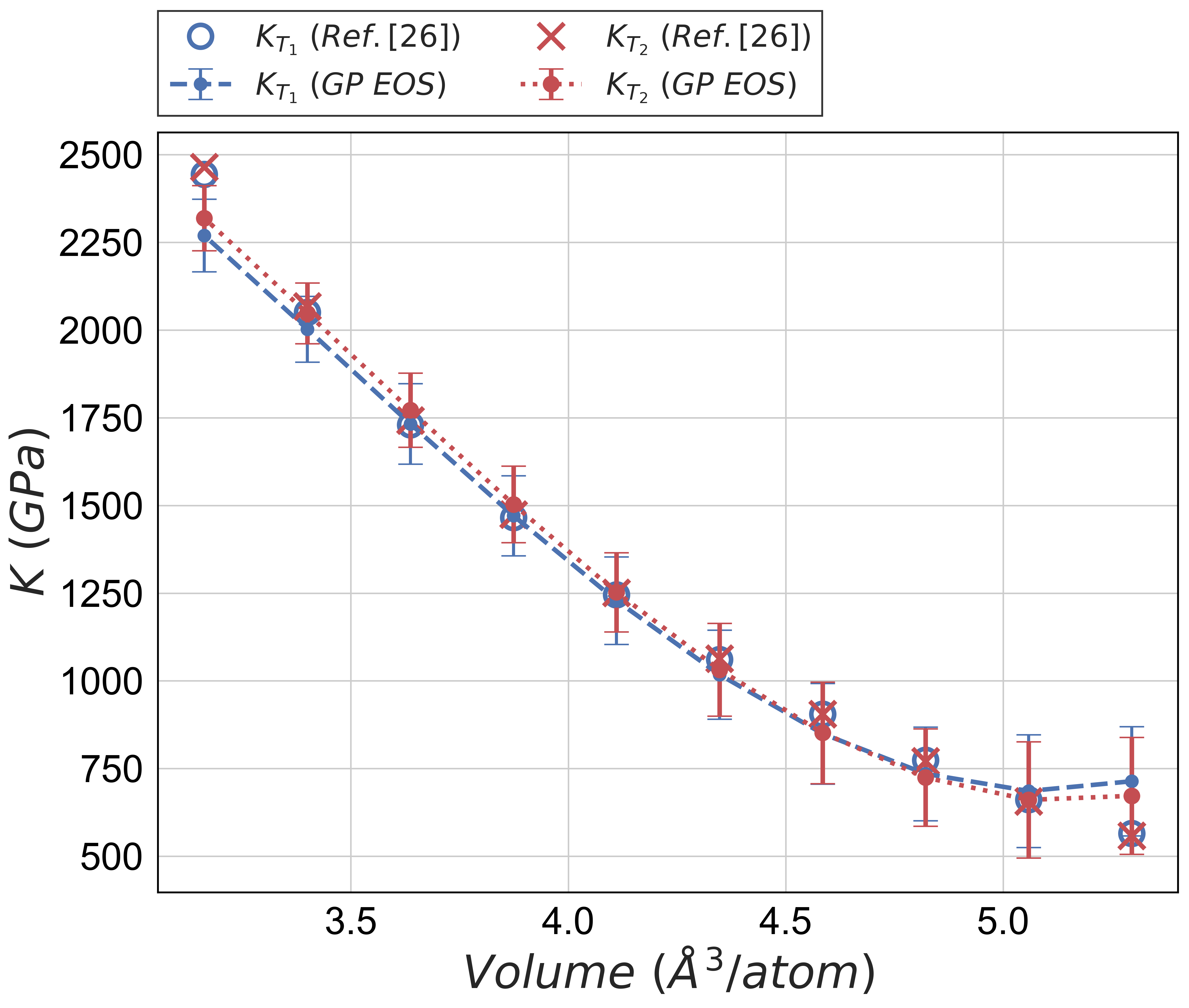} &
\includegraphics[width=0.48\textwidth]{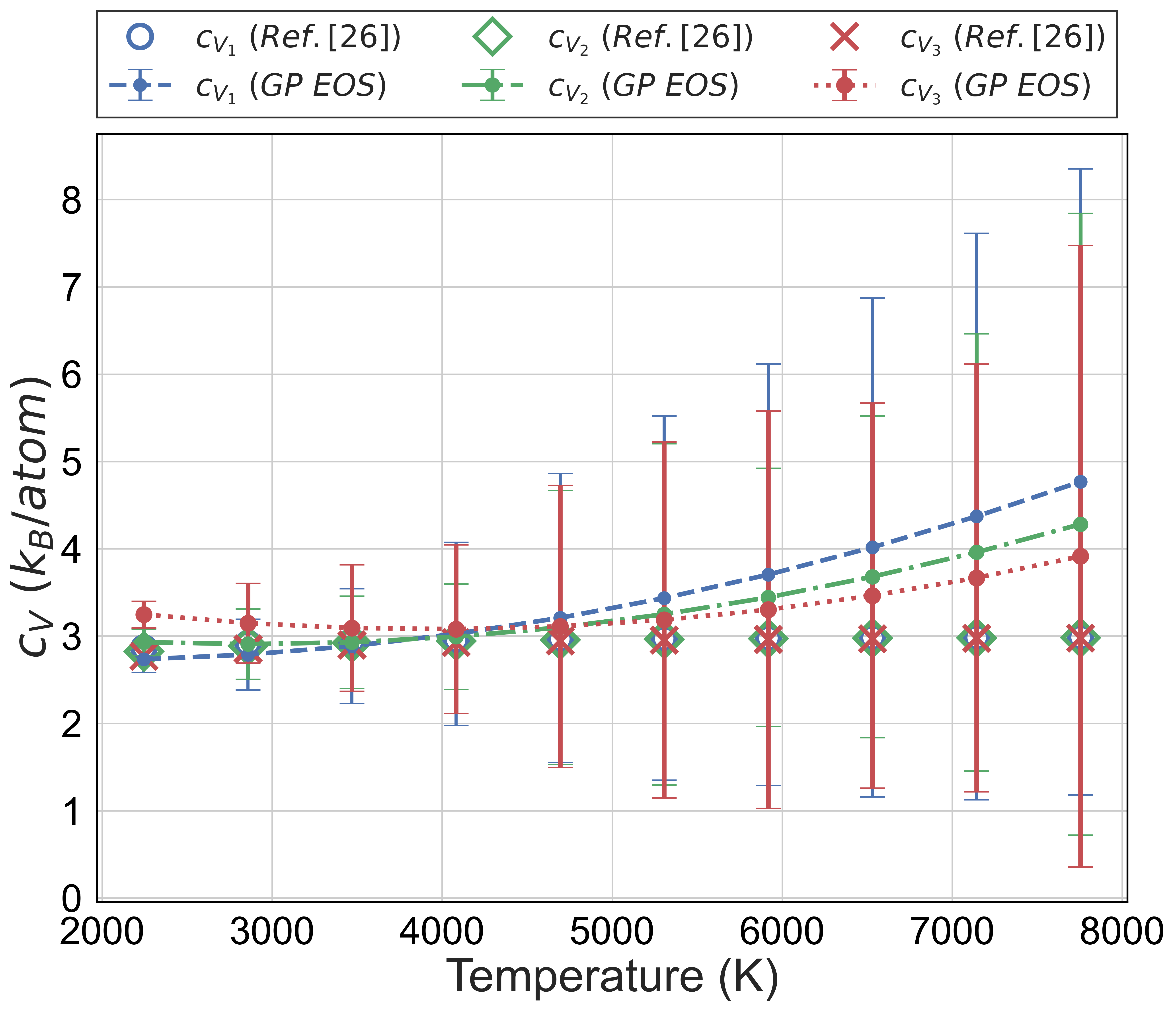} \\ 
(c)& (d)\\
\end{tabular}
\caption{Plots of the validation study between the proposed GP EOS model and parametric Benedict EOS model:\cite{benedict2014multiphase} (a) Percentage error$^{*}$ between mean pressure GP EOS and Benedict pressure EOS.  (b) Percentage error$^{*}$ between mean energy GP EOS and Benedict energy EOS. (c) Comparison of isothermal bulk modulus ($K_T$) obtained from GP EOS model (with $95\%$ confidence bounds) and Benedict EOS\cite{benedict2014multiphase} for $T_1 = 2490K$ and $T_2=7510K$. (d) Comparison of specific isochoric heat capacity ($c_V$) obtained from GP EOS model (with $95\%$ confidence bounds) and Benedict EOS\cite{benedict2014multiphase} for $V_1=4.8\ A^3/atom$, $V_2=4.2\ A^3/atom$ and $V_3=3.6\ A^3/atom$. *$\Big|\frac{X_{\text{mean GP EOS}}-X_{\text{Benedict EOS}}}{X_{\text{Benedict EOS}}}\Big|\times 100$ }
\label{fig:validation_plots}
\end{figure}
These results are supported by plotting the thermodynamic stability conditions in the $(V,T)$ space in 
Figures \ref{fig:constraint_plots}(a) and (b) for the constrained P-GP and E-GP, respectively. We see that the thermodynamic stability constraints are satisfied across the domain in the proposed constrained EOS model. Meanwhile, Figures \ref{fig:constraint_plots}(c) and (d) show the stability constraints for the unconstrained P-GP and E-GP EOS models, respectively, where we see that the energy stability constraint (Eq.\ \eqref{stability2}) is violated in regions characterized by high temperatures across volumes within the domain.. This results in a negative specific heat in these regions, which will likely cause a hydrodynamics simulation to crash. The pressure stability constraint, on the other hand, is not violated even in the unconstrained GP. However, we note that the unconstrained GP EOS does not satisfy the thermodynamic consistency constraint because the covariance model is not physics informed. 

Figure \ref{fig:constraint_plots} further supports the observations made in Figure \ref{fig:constrainedPE-GP}. 
The reduction in uncertainty in the constrained E-GP (Figure \ref{fig:constrainedPE-GP}(b)) compared to unconstrained E-GP (Figure \ref{fig:constrainedPE-GP}(d)) is due to the introduction of the energy stability constraint. 
Moreover, since the proposed GP has a physics-informed covariance model, which yields valid cross-covariances between the pressure and energy GP, improvement in the prediction of E-GP also improves predictions of P-GP even though the pressure stability constraints are not violated in the unconstrained P-GP.
Hence, the resulting physics-informed data-driven EOS model provides an excellent fit to the DFT-MD data and a realistic estimate of total prediction uncertainty. 

However, we note that the simulation data used for training do not include estimates of uncertainty. If such data uncertainty were provided, they could be easily incorporated into the proposed framework through the training described in Section \ref{Section 3}.  

We have validated the results obtained from our data driven GP EOS model with the parameteric Benedict EOS\cite{benedict2014multiphase} in Figure \ref{fig:validation_plots}. 
In Figures \ref{fig:validation_plots}(a) and (b), we first plotted the percentage error for pressure and energy between the ones predicted by the Benedict EOS \cite{benedict2014multiphase} and the mean of our GP EOS. As can be observed from the Figure \ref{fig:validation_plots}(a), the percentage error in pressure remains relatively low across most regions in the $(V,\ T)$ space. In specific regions characterized by high volume and low temperature, as well as high volume and temperature, there is a moderate deviation, typically ranging between $5-10\%$. For these regions, our pressure GP EOS also gives the highest uncertainty. Similarly, in Figure \ref{fig:validation_plots}(b), the percentage error in energy is relatively low across most $(V,\ T)$ regions with a moderate deviation (around $3-5.5\%$) for high volume and temperature. Again, our energy GP EOS gives the highest uncertainty for this region. This indicates the effectiveness of our model in accurately representing the characteristic uncertainty in the model.

We also obtained the specific isochoric heat capacity  ($c_V$) and isothermal bulk modulus ($K_T$) from the Benedict EOS \cite{benedict2014multiphase} and the proposed GP EOS model. These quantities, which were not explicitly included in the training of our model but are derived from it by numerical differentiation, serve to further validate the GP EOS and the uncertainties from it. In Figure \ref{fig:validation_plots}(c), we plot the isothermal Bulk modulus against volume. The figure shows that the mean $K_T$ obtained from our model aligns well with the values derived from the Benedict EOS \cite{benedict2014multiphase}. Any observed deviation is effectively characterized within the $95\%$ confidence bounds. Additionally, it can be seen that $K_T$ does not vary appreciably with temperature within the specified temperature range.

Similarly, for $c_V$ we can observe good agreement between the mean $c_V$ obtained from our model and Benedict EOS in Figure~\ref{fig:validation_plots}(d). As pointed out in the literature~\cite{benedict2014multiphase, correa2008first}, surprisingly, the specific heat is observed to be indistinguishable from $3k_B/atom$ up to 20,000 K and is also very nearly independent of volume. Our model captures these observations within the 95\% confidence intervals, but suggests there is significant uncertainty in $c_V$ particularly at high temperature. This uncertainty results from seemingly small uncertainties in the EOS that grow considerably when differentiated to compute $c_V=\dfrac{\partial E}{\partial T}\biggr|_V$.






\subsection{GP Hugoniot Curve for Diamond}

\begin{figure}[!ht]
   \centering 
    \begin{tabular}{cc}
        \includegraphics[width=0.45\textwidth]{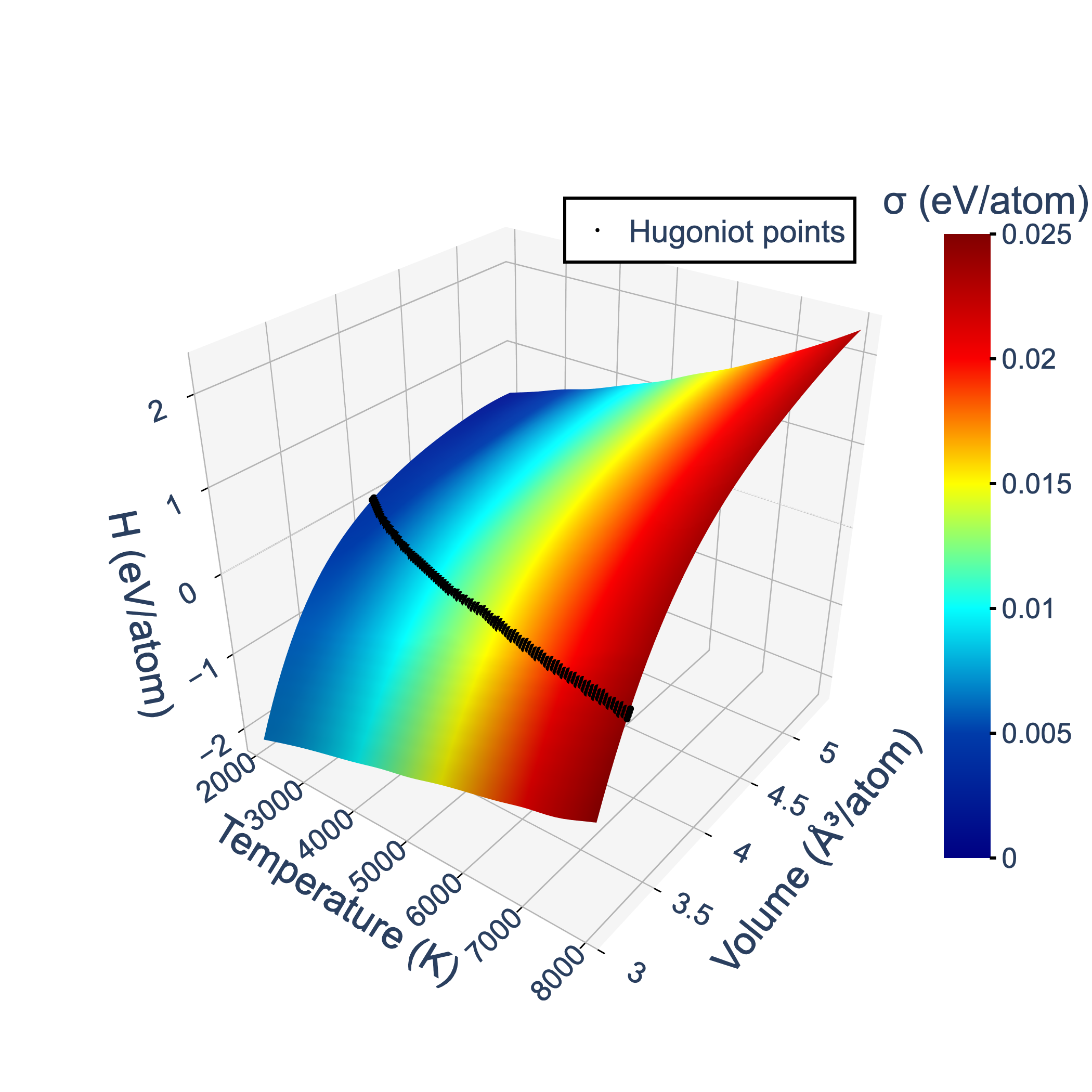} &
        \raisebox{5mm}{\includegraphics[width=0.4\textwidth]{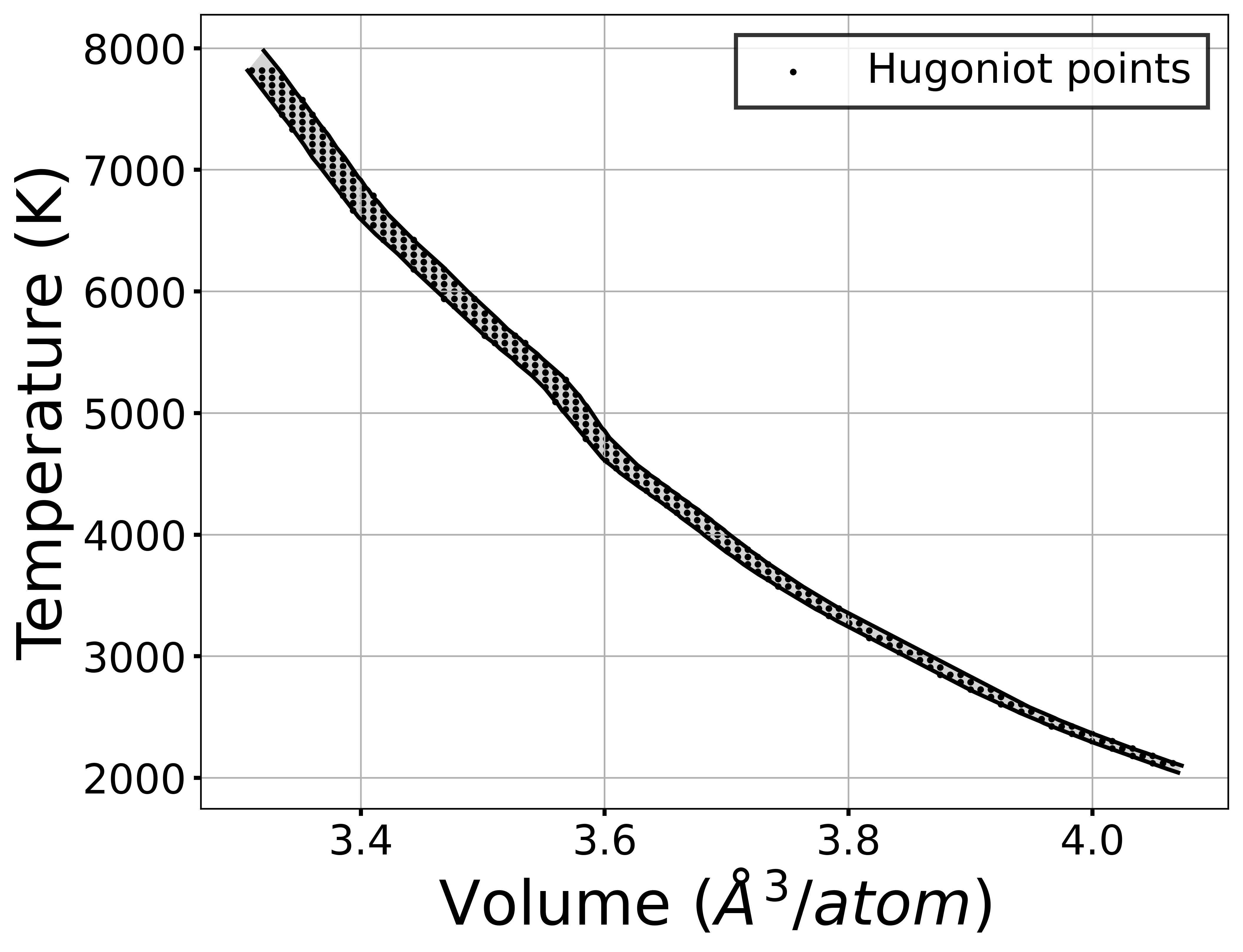}} \\
        (a) & (b) \\
    \end{tabular}
    \caption{(a) GP Hugoniot function, $H(V,T)$ showing points where $H=0$ lies within the 95\% confidence interval of the GP in black. (b) Hugoniot points where $H=0$ lies within the 95\% confidence interval of the GP shown in the $(V,T)$ plane.}
    \label{fig:Hugoniot}
\end{figure}

\begin{figure}[!ht]%
 \centering
 \subfloat[]{\includegraphics[width=0.5\textwidth]{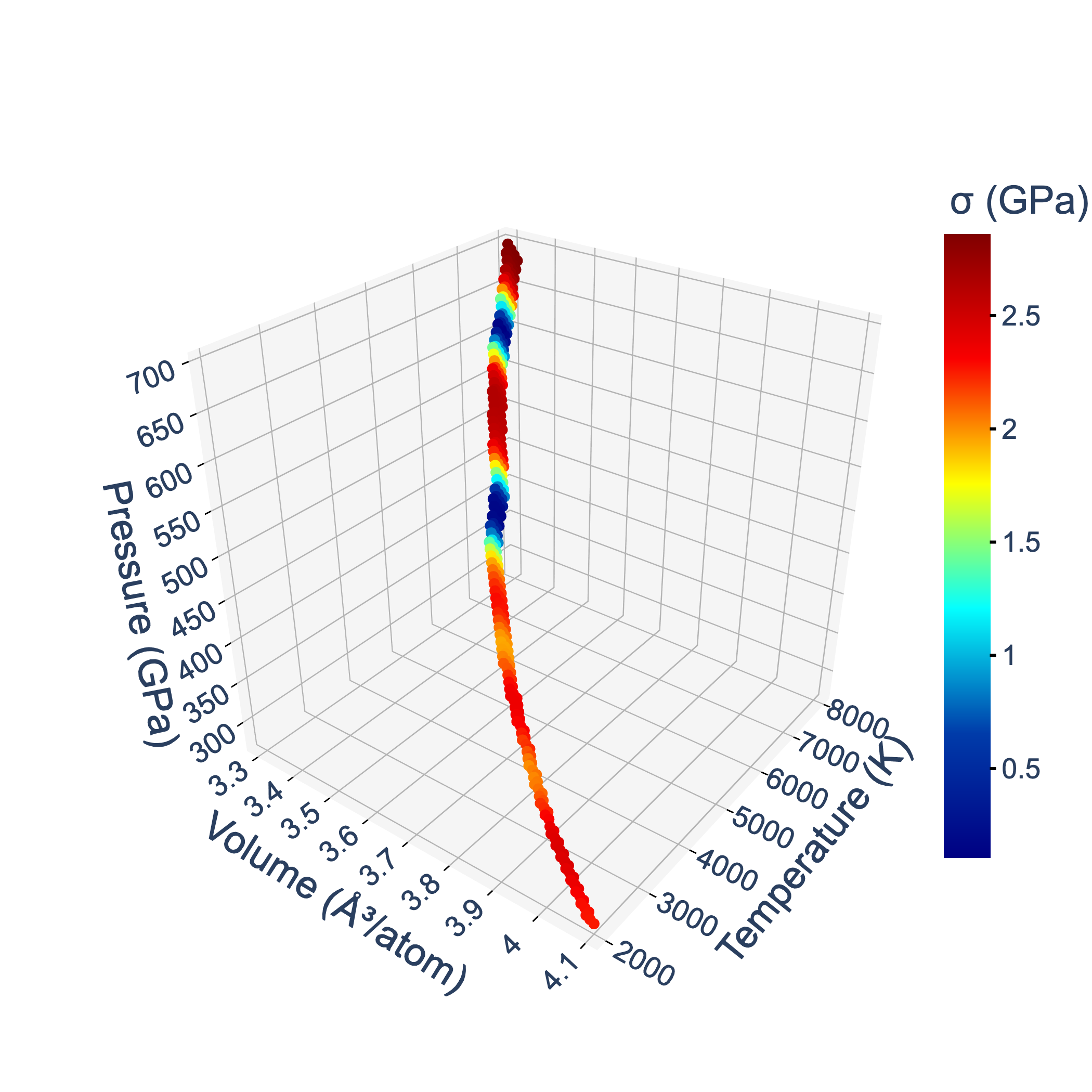}\label{fig:a}}\\
 \subfloat[]{\includegraphics[width=0.45\textwidth]{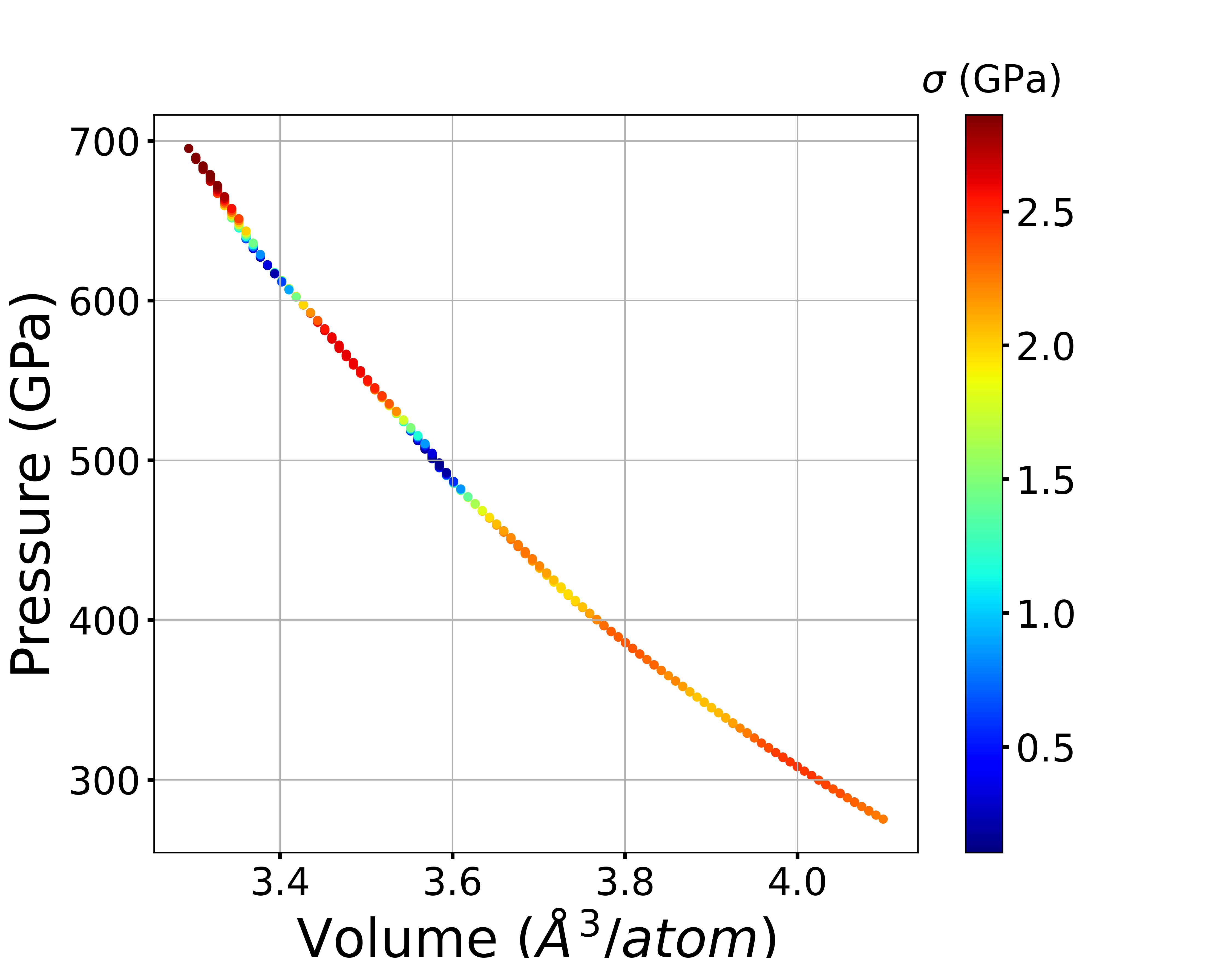}\label{fig:b}}
 \subfloat[]{\includegraphics[width=0.45\textwidth]{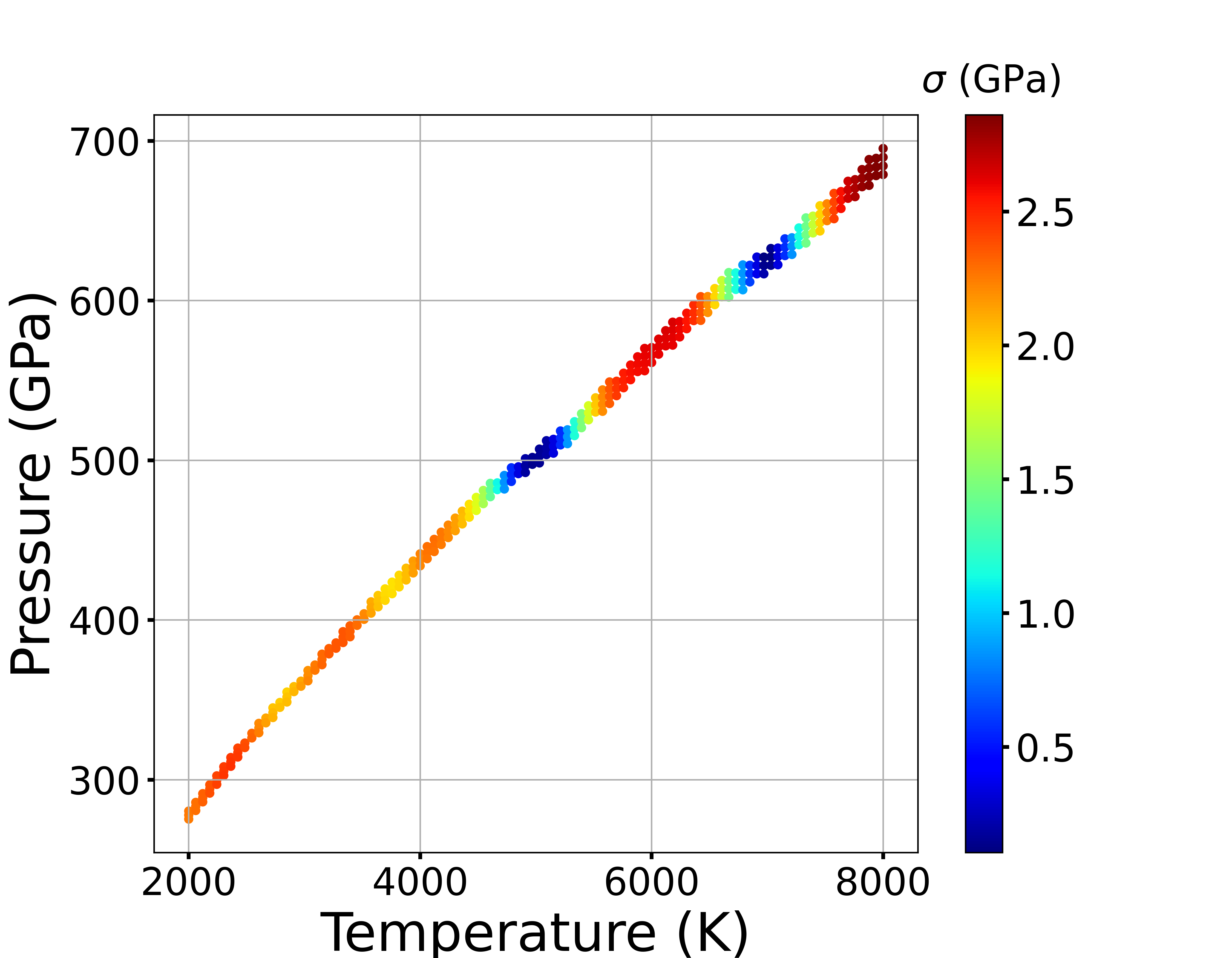}\label{fig:c}}%
 \caption{Plots of the uncertain Hugoniot curve. The width of the curve represents uncertainty in the position of the Hugoniot (i.e. satisfying $H=0$ with 95\% confidence) along the EOS and the corresponding uncertainty in pressure is shown with colors denoting the standard deviation. (a) Complete Hugoniot in $(V,T,P)$ space showing all points along the EOS in which $H=0$ lies within the 95\% confidence intervals of the Hugoniot GP. (b) Uncertain Hugoniot curve as a function of Volume. (c) Uncertain Hugoniot curve as a function of Temperature.}%
 \label{fig:uncertain_Hugoniots}%
\end{figure}

The GP Hugoniot function, $H(V,T)$ is derived from the physics-constrained GP EOS model following the formulation presented in Section \ref{B} and illustrated in Figure \ref{fig:Hugoniot}. 
The input states $(V, T)$ that satisfy Eq.\ \eqref{hugo} for which $H =0$ lies in the $95\%$ confidence interval of H-GP are also shown in Figure \ref{fig:Hugoniot}(a). These Hugoniot points are plotted separately in Figure \ref{fig:Hugoniot}(b), and a deterministic curve is fit to establish a mapping $V_H\rightarrow T_H$. For these Hugoniot points $(V_H, T_H)$, the predictive distribution of pressure and energy is computed using Eq.\ \eqref{eqns:PE_pred}. The resulting Hugoniot points are shown in Figure \ref{fig:uncertain_Hugoniots}(a) where uncertainties in pressure are shown with color representing the standard deviation. The corresponding projections of the Hugoniot showing the P-V and P-T curves are shown in Figure \ref{fig:uncertain_Hugoniots}(b) and (c), respectively. Thus, we have demonstrated a way to derive the Hugoniot with uncertainty directly from the physics-constrained GP EOS. 

\subsection{Unified GP EOS for Diamond Trained from Simulations and Experiments}

\begin{figure}[!ht]
   \centering 
    \begin{tabular}{cc}
\includegraphics[width=0.45\textwidth]{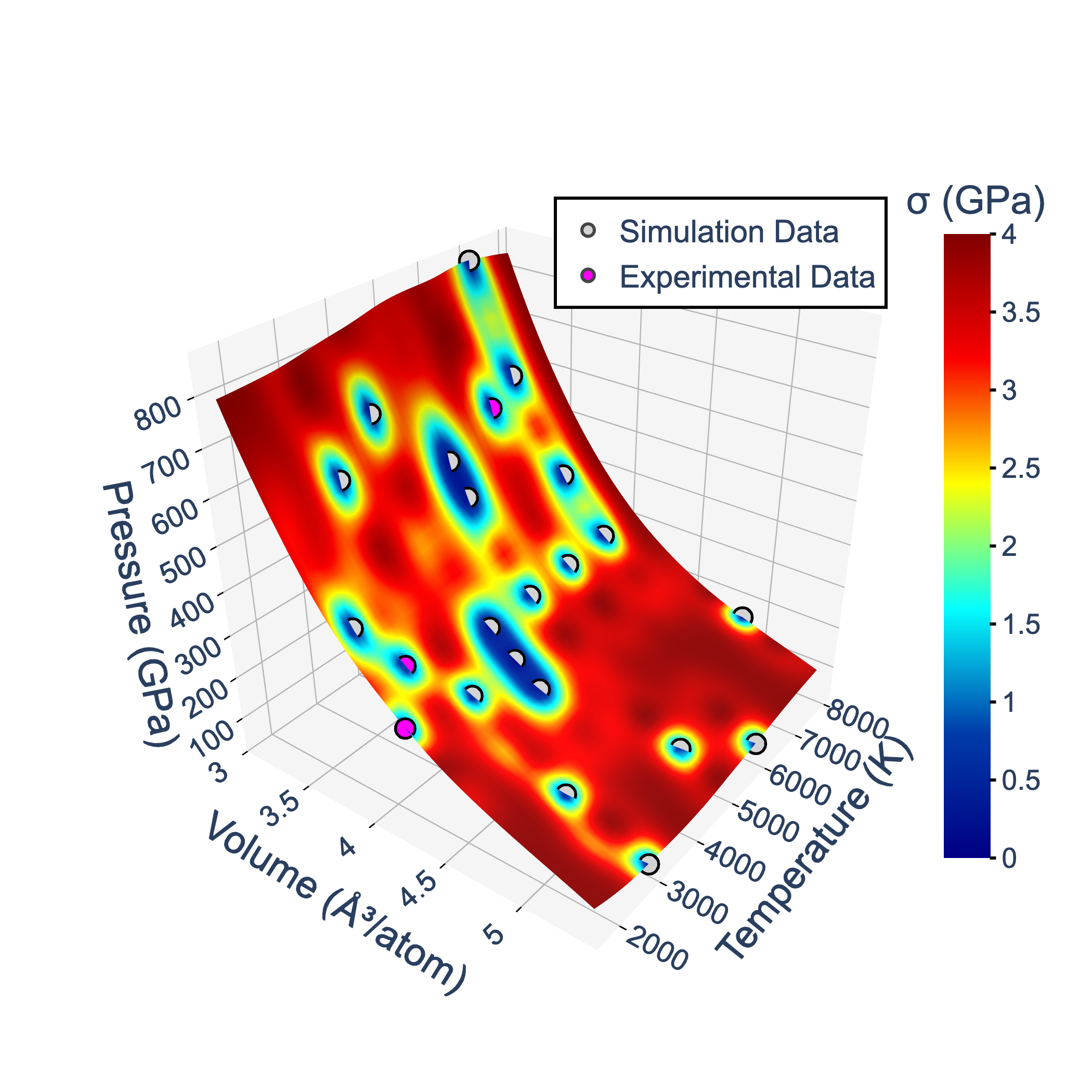} &
\includegraphics[width=0.45\textwidth]{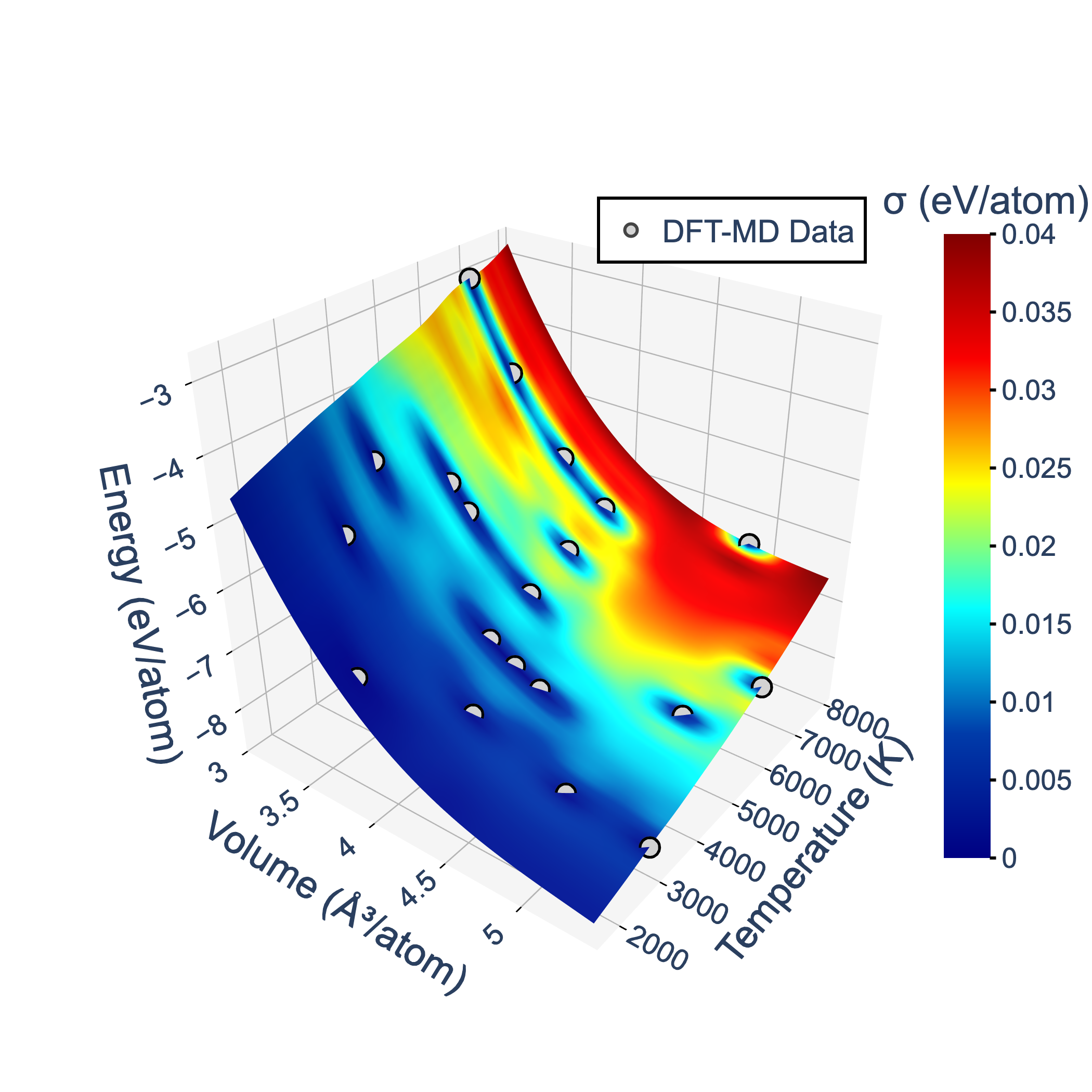} \\ 
(a)& (b)\\
\end{tabular}
\caption{Unified constrained GP EOS model trained on 20 DFT-MD simulation data and 3 shock compression experimental data relating pressure and volume. (a) Pressure marginal GP EOS model (b) Energy marginal GP model}
\label{fig:unifiedGP}
\end{figure}

\begin{figure}[h!]
   \centering 
    \begin{tabular}{cc}
\includegraphics[width=0.45\textwidth]{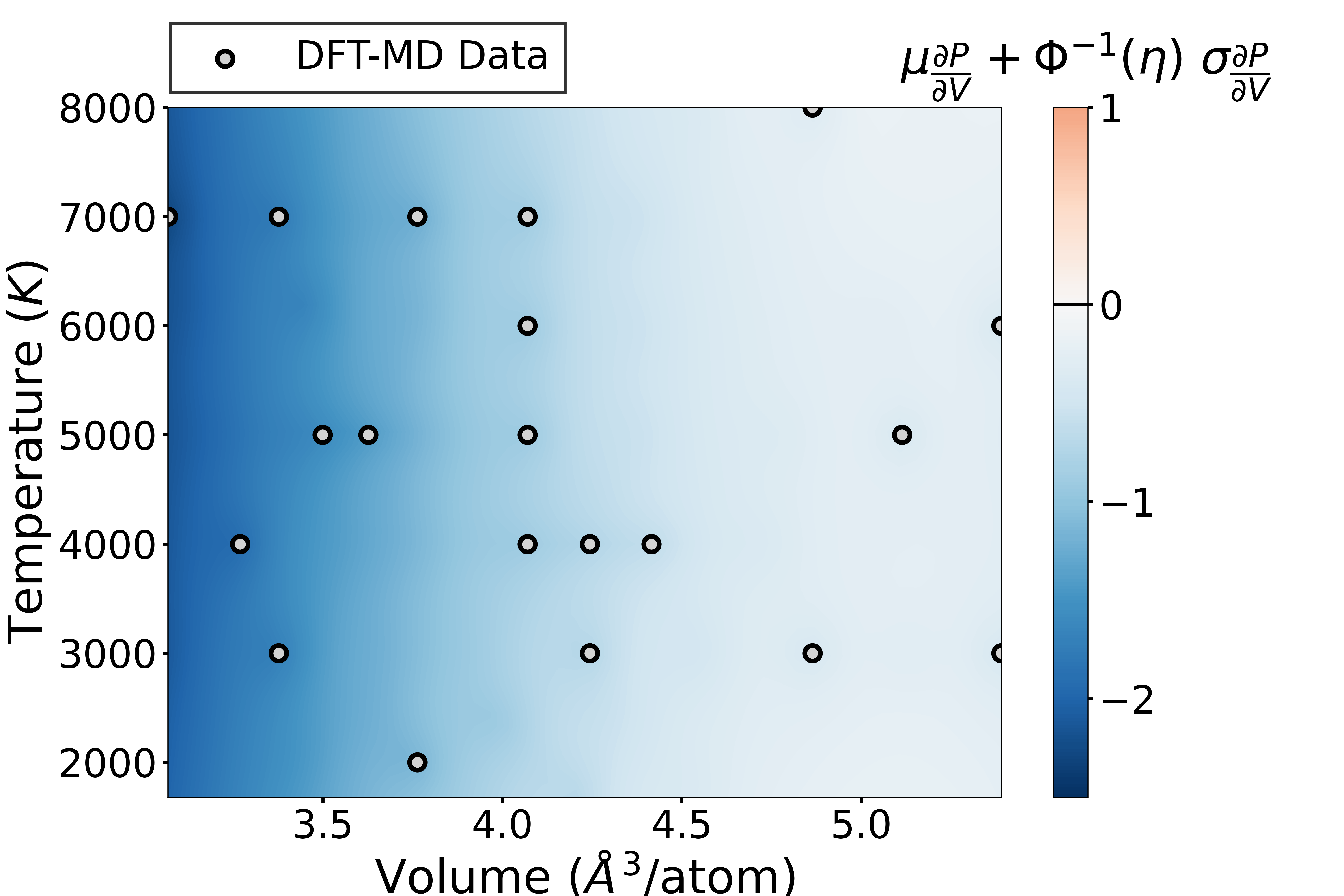} &
\includegraphics[width=0.45\textwidth]{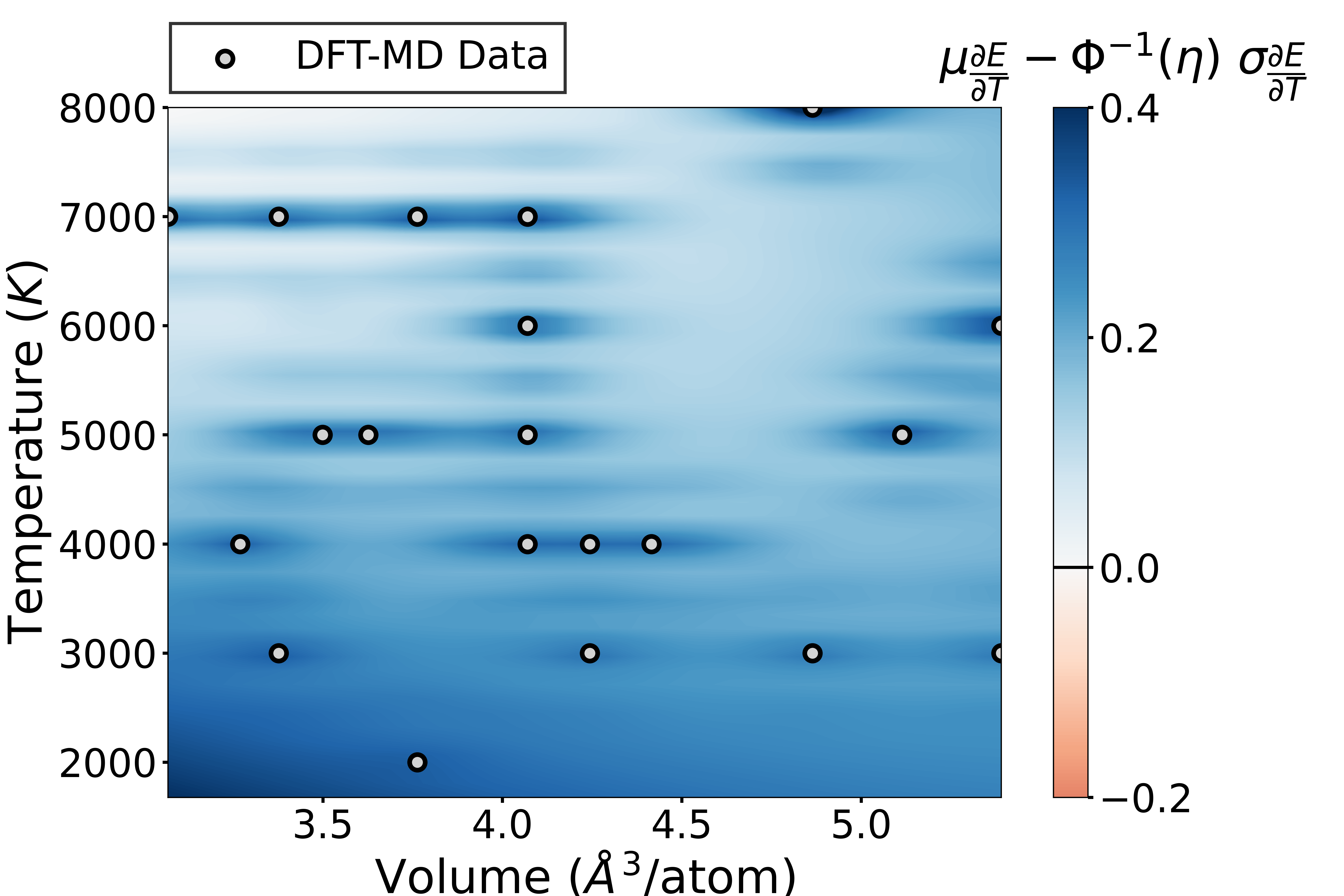} \\ 
(a)& (b)\\
\end{tabular}
\caption{Plots of the thermodynamic stability conditions in the $(V,T)$ space for the unified constrained GP EOS trained on simulation and experimental data. Regions of constraint violation are represented by red. (a) Probabilistic stability constraint for unified constrained Pressure GP EOS model showing that the constraints are satisfied. (b) Probabilistic stability constraint for unified constrained Energy GP EOS model showing that the constraints are satisfied.}
\label{fig:unifiedGP_stability}
\end{figure}

In this section, we present a unified physics-informed GP EOS for the diamond phase of carbon that is trained from multiple data sources and provides accurate EOS predictions with uncertainty -- as described in Section \ref{C}. The given DFT-MD simulation data (same 20 data points used above) is augmented by Hugoniot experimental data from dynamic shock wave experiments that relate pressure and volume\cite{mcwilliams2010strength} (3 points). These data are from a pressure regime where the diamond has significant strength, meaning that we expect to see a discrepancy between simulation and experiment which provides an interesting test for our unified modeling approach. Additionally, the temperature is not usually measured in these high-pressure experiments (except in those with static compression), which provides another difficulty to overcome with our approach. The GP EOS model trained in the previous sections can be used to estimate temperature for a given volume using the mapping  ($V_H\rightarrow T_H$) shown in Figure \ref{fig:Hugoniot}(b). We apply a deterministic mapping here, but theoretically, we could also construct a GP relating temperature and volume, which would formally provide a probability distribution for temperature. However, such an approach requires a more challenging deep GP framework (since the input state $T$ is a GP), which is beyond the scope of the present work. 
With these temperature estimates, we now have a training set $\{\mathbf{X}_P, P$\} (23 data points) and $\{\mathbf{X}_E, E\}$ (20 data points). We use this training data to train the unified GP. The results are shown in Figure \ref{fig:unifiedGP}. As expected, the uncertainty reduces significantly near the experimental observations in Figure \ref{fig:unifiedGP}(a). 
Comparing Figure \ref{fig:unifiedGP}(a) with Figure \ref{fig:constrainedPE-GP}(a), we can observe an increase in the uncertainty, which captures the inconsistencies between the simulation and experimental data. Since the P-GP and E-GP are correlated, a slight increase in uncertainty is also observed in Figure \ref{fig:unifiedGP}(b) compared to \ref{fig:constrainedPE-GP}(b).
Nonetheless, the overall uncertainty is still  relatively low in terms of COV. However, if the data is highly inconsistent or a different interpretation of the difference between model and experiment is desired, various approaches from the literature can be considered to address the inconsistency. For example specific analysis may choose a suitable kernel (e.g., Matérn kernel with higher smoothness parameters), more robust likelihood functions (e.g., Huber likelihood), regularizations, and cross-validations among others\cite{williams2006gaussian}, while still ensuring physical consistency in the manner we have described. Finally, as shown in Figure \ref{fig:unifiedGP_stability}, the unified GP EOS still satisfies the thermodynamic stability constraints as necessary. We have therefore developed a comprehensive unified framework that has been successfully trained using both first-principles simulation data and experimental shock compression experiments.   

\section{Conclusion}

In this work, we have developed a novel data-driven framework to construct thermodynamically constrained equation of state (EOS) models with uncertainty. The proposed framework is based on a non-parametric constrained Gaussian process regression (GPR), which inherently captures the model and data uncertainties while satisfying the essential thermodynamic stability and consistency constraints. Violation of these constraints results in a non-physical EOS that will cause problems in downstream applications such as hydrodynamics simulation. The key benefits of using GPR to build the EOS model is that it can be trained on relatively small data sets compared to other machine learning methods like neural networks and automatically estimates the prediction uncertainty. The resulting EOS model also yields a GP for the shock Hugoniot with uncertainty, which has been derived herein. Further, we proposed a unified framework such that the GP can leverage both simulation and experimental data and provides pointwise EOS predictions with uncertainty. The resulting EOS can therefore be directly incorporated into hydrocode simulations for uncertainty quantification studies. 

We have specifically demonstrated the training of this physics-constrained GP EOS for the diamond phase of Carbon from first-principles DFT-MD simulations. We show that the model satisfies the thermodynamic constraints, which results in a reduction in uncertainty at certain points as well. In short, we show that considering thermodynamic constraints improves confidence in EOS predictions and supplements limited data. We then derive the Hugoniot for diamond from the GP EOS and demonstrate that the trained model can be augmented with experimental shock Hugoniot data to improve the EOS -- thus demonstrating our unified framework for EOS training. 

The proposed framework can be similarly applied to different material phases. However, an extension of the proposed framework to a more generalized multiphase EOS model that captures phase transitions is the subject of further work. Finally, we anticipate that the proposed framework opens the door to a wide range of improvements in EOS modeling and the associated downstream applications. For example, in future studies the prediction uncertainty can be used to inform the choice of points for new simulations/experiments through e.g.\ Bayesian optimization resulting in smaller data sets requirements and thus accelerating the development of new EOSs. Moreover, the proposed framework can be integrated with physics-based parametric models similar
to the KOH approach\cite{kennedy2001bayesian}, such that the physically constrained GP serves as a correction to the parametric model and potentially facilitates multi-fidelity modeling. Finally, these physics-informed GP EOS models will be integrated into hydrocode simulations of shock experiments to enable uncertainty quantification studies.  


\section*{Data Availability Statement}
Code developed in this work, training data, and presented results are available through the following Github repository: \url{https://github.com/himanshu258/constrained_GP_EOS}~\cite{Github}.

\begin{acknowledgments}
This work was performed under the auspices of the U.S. Department of Energy by Lawrence Livermore
National Laboratory under Contract DE-AC52-07NA27344, and was supported by Defense Threat Reduction Agency, Award HDTRA12020001. LLNL-JRNL-850088
\end{acknowledgments}

\bibliography{main}
\appendix
\renewcommand{\theequation}{A.\arabic{equation}}
\setcounter{equation}{0}
\section*{APPENDIX}
The Hugoniot equation can be expressed as,
\begin{equation}
H(V,T)=E(V,T)-E_0+\frac{1}{2}\left(V-V_0\right)\left(P(V,T)+P_0\right)
\end{equation}
where $E_0$, $P_0$ and $V_0$ are initial energy, pressure and volume, respectively. 

We recognize that $H = f\left(P, E\right)$, where $f$ is a linear function of both $P$ and $E$. Applying the Taylor series expansion about its mean, we get
\begin{equation}
   H = f\left(P, E\right)=f\left(\mu_{P}, \mu_{E} \right)+ \left.\frac{\partial f\left(P, E\right)}{\partial P}\right|_{\mu_{P}, \mu_{E}} \left(P-\mu_{P}\right) + \left.\frac{\partial f\left(P, E\right)}{\partial E}\right|_{\mu_{P}, \mu_{E}} \left(E-\mu_{E}\right)
   \label{eqns:hugoniot}
\end{equation}
Since both $P$ and $E$ are GPs, $H$ (a linear function of $P$ and $E$) is also a GP. Taking the expectation of Eq.\ \eqref{eqns:hugoniot} yields:
\begin{equation}
   \mathop{{}\mathbb{E}}\left[H\right] = \mathop{{}\mathbb{E}}\left[f\left(P, E\right)\right]=f\left(\mu_{P}, \mu_{E} \right) = \mu_{H}
   \label{eqns:meanH}
\end{equation}
The covariance of the $H$ GP is obtained by:
\begin{equation}
\small
\begin{aligned}
   \mathop{{}\mathbb{E}}\Big[(H-\mathop{{}\mathbb{E}}[H])(H-\mathop{{}\mathbb{E}}[H])\Big] 
   & = \mathop{{}\mathbb{E}}\bigg[\bigg(\left.\frac{\partial f\left(P, E\right)}{\partial P}\right|_{\mu_{P}, \mu_{E}} \left(P-\mu_{P}\right) + \left.\frac{\partial f\left(P, E\right)}{\partial E}\right|_{\mu_{P}, \mu_{E}} \left(E-\mu_{E}\right)\bigg)\\
   &\qquad \times\bigg( \bigg(\left.\frac{\partial f\left(P, E\right)}{\partial P}\right|_{\mu_{P}, \mu_{E}} \left(P-\mu_{P}\right) + \left.\frac{\partial f\left(P, E\right)}{\partial E}\right|_{\mu_{P}, \mu_{E}} \left(E-\mu_{E}\right)\bigg)\bigg]\\
   & = \bigg(\left.\frac{\partial f\left(P, E\right)}{\partial P}\right|_{\mu_{P}, \mu_{E}}\bigg)^{2}\mathop{{}\mathbb{E}}\bigg[\left(P-\mu_{P}\right) \left(P-\mu_{P}\right)\bigg]  +  \bigg(\left.\frac{\partial f\left(P, E\right)}{\partial E}\right|_{\mu_{P}, \mu_{E}}\bigg)^{2} \mathop{{}\mathbb{E}}\bigg[\left(E-\mu_{E}\right) \left(E-\mu_{E}\right)\bigg]\\ & \qquad+ 2\bigg(\left.\frac{\partial f\left(P, E\right)}{\partial E}\right|_{\mu_{P}, \mu_{E}}\bigg)\bigg(\left.\frac{\partial f\left(P, E\right)}{\partial P}\right|_{\mu_{P}, \mu_{E}}\bigg) \mathbb{E}\bigg[\left(E-\mu_{E}\right) \left(P-\mu_{P}\right)\bigg] \\  
   & = \frac{1}{4}\left(V-V_0\right)^{2} \operatorname{Cov}\left(P, P\right) + \operatorname{Cov}\left(E, E\right) + (V-V_0)\operatorname{Cov}(E,P) \\
K_{HH}\left(\mathbf{X}, \mathbf{X}^{\prime}\right)   & = \frac{1}{4}\left(V-V_0\right)^{2} K_{PP}\left(\mathbf{X}, \mathbf{X}^{\prime}\right)+ K_{EE}\left(\mathbf{X}, \mathbf{X}^{\prime}\right) + (V-V_0)K_{PE}\left(\mathbf{X}, \mathbf{X}^{\prime}\right)
\end{aligned}
\label{eqns:covHH}
\end{equation}
The cross-covariance of the $H$ and $P$ GPs is given by
\begin{equation}
\small
\begin{aligned}
   \mathop{{}\mathbb{E}}\Big[(H-\mathop{{}\mathbb{E}}[H])(P-\mathop{{}\mathbb{E}}[P])\Big] &= \mathop{{}\mathbb{E}}\bigg[\bigg( \left.\frac{\partial f\left(P, E\right)}{\partial P}\right|_{\mu_{P}, \mu_{E}} \left(P-\mu_{P}\right) + \left.\frac{\partial f\left(P, E\right)}{\partial E}\right|_{\mu_{P}, \mu_{E}} \left(E-\mu_{E}\right) \bigg)\bigg( \left(P-\mu_{P}\right)  \bigg)\bigg]\\
   & = \left.\frac{\partial f\left(P, E\right)}{\partial P}\right|_{\mu_{P}, \mu_{E}}\mathop{{}\mathbb{E}}\bigg[\left(P-\mu_{P}\right) \left(P-\mu_{P}\right)\bigg]  + \left.\frac{\partial f\left(P, E\right)}{\partial E}\right|_{\mu_{P}, \mu_{E}} \mathop{{}\mathbb{E}}\bigg[\left(E-\mu_{E}\right) \left(P-\mu_{P}\right)\bigg] \\  
   & = \frac{1}{2}\left(V-V_0\right) \operatorname{Cov}\left(P, P\right) + \operatorname{Cov}\left(E, P\right) \\
  K_{HP}\left(\mathbf{X}, \mathbf{X}^{\prime}\right) & = \frac{1}{2}\left(V-V_0\right) K_{PP}\left(\mathbf{X}, \mathbf{X}^{\prime}\right) + K_{EP}\left(\mathbf{X}, \mathbf{X}^{\prime}\right)  
\label{eqns:K_HP}
\end{aligned}
\end{equation}
Similarly, we can obtain the cross-covariance between the $H$ and $E$ GPs as, 
\begin{equation}
    K_{HE}\left(\mathbf{X}, \mathbf{X}^{\prime}\right)  = K_{EE}\left(\mathbf{X}, \mathbf{X}^{\prime}\right)+ K_{EP}\left(\mathbf{X}, \mathbf{X}^{\prime}\right)
\label{eqns:K_EH}
\end{equation}
For cases where the GP EOS is intended for applications where a linear relationship may not exist, we can still derive the mean and covariance model in the same manner by considering higher-order closure terms in the Taylor series approximation. \citet{constantinescu2013physics}  demonstrated how to address non-linear dependencies between GP outputs for physics-based systems to obtain a valid covariances model. There are also recent works in the multioutput GP literature that deals with obtaining covariance models for systems that are non-linear while ensuring positive definiteness\cite{hu2021nonlinear, alvarez2019non}.
\end{document}